\documentclass[12pt]{iopart}
\usepackage{graphicx}
\usepackage{lineno}
\usepackage{amssymb}

\begin{document}

%\linenumbers

%\title[Author guidelines for IOP Publishing journals in  \LaTeXe]{How to prepare and submit an article for publication in an IOP Publishing journal using \LaTeXe}

\title[Cumulants and ordering of their ratios in two-dimensional Potts models]{Finite-size behavior of higher-order cumulant ratios near criticality in two-dimensional Potts models}

\author{Rajiv V. Gavai$^{1}$, Bedangadas Mohanty$^{2,3}$, Jaydev Singh Rao$^{4}$ and Swati Saha$^{2,3,*}$}                     % Do not remove

\address{
  $^{1}$ Department of Physics, Indian Institute of Science Education and Research Bhopal, Bhopal 462066, Madhya Pradesh, India \\
  $^{2}$ School of Physical Sciences, National Institute of Science Education and Research, Jatni 752050, Odisha, India \\
  $^{3}$ Homi Bhabha National Institute, Training School Complex, Anushaktinagar, Mumbai 400094, Maharastra, India \\
  $^{4}$ Department of Electrical Engineering and Computer Science, Indian Institute of Science Education and Research Bhopal, Bhopal 462066, Madhya Pradesh, India\\
  $^{*}$ Author to whom any correspondence should be addressed}

\ead{gavai@tifr.res.in, bedanga@niser.ac.in, jaydevsrao@gmail.com, and swati.saha@niser.ac.in}
%\ead{swati.saha@niser.ac.in}

\vspace{10pt}
\begin{indented}
\item[]February, 2026
\end{indented}

\begin{abstract}
Theoretical considerations predict a specific hierarchy among ratios of net-baryon number cumulants ($\chi_n$, where $n$ is the order of cumulant) in the vicinity of the transition from the low-temperature hadronic phase to the high temperature quark-gluon plasma phase at small baryon chemical potential, $\mu_\mathrm{B}$, in the QCD phase diagram. This hierarchy, $\frac{\chi_6}{\chi_2} < \frac{\chi_5}{\chi_1} < \frac{\chi_4}{\chi_2} < \frac{\chi_3}{\chi_1}$, has been observed by the STAR experiment in net-proton number (a proxy of net-baryon number) cumulant ratios over a broad range of collision energies. Motivated by these findings, we investigate whether similar ordering emerges generically in finite statistical systems undergoing second-order phase transitions. 

We employ two different spin models: the two-state and three-state Potts models in two dimensions, both exhibiting a transition from an ordered phase to a disordered phase at their respective critical temperatures. Monte Carlo simulations are performed on square lattices of varying sizes using the Wolff cluster algorithm. Cumulants of the total magnetization are calculated up to sixth order in both of these models in a temperature range near their corresponding critical temperatures. Higher-order cumulants exhibit extrema (peaks/troughs) whose magnitudes grow with both cumulant order and lattice size, reflecting enhanced critical fluctuations. Except within a narrow temperature window above the critical temperature, neither the complete hierarchy nor its exact reverse is realized over the studied temperature range in either model.

\end{abstract}

%
% Uncomment for keywords
%\vspace{2pc}
%\noindent{\it Keywords}: XXXXXX, YYYYYYYY, ZZZZZZZZZ
%
% Uncomment for Submitted to journal title message
%\submitto{\JPA}
%
% Uncomment if a separate title page is required
%\maketitle
% 
% For two-column output uncomment the next line and choose [10pt] rather than [12pt] in the \documentclass declaration
%\ioptwocol
%

\section{\label{sec:Introduction}Introduction}
Understanding the phase structure and the nature of phase transitions in the temperature-baryon density (or equivalently
the baryon chemical potential, $\mu_\mathrm{B}$) plane for the strongly interacting matter, known as quantum chromodynamics (QCD) phase diagram, is crucial for unraveling the fundamental properties of matter \cite{Shuryak:1980tp,Ding:2015ona}. While this diagram is not well-explored for larger $\mu_\mathrm{B}$, the region below $\mu_\mathrm{B}$ = 450 MeV ($\mu_\mathrm{B}/T_\mathrm{ch} \sim$ 2.8--3, where $T_\mathrm{ch}$ is the chemical freeze-out temperature inferred from particle yields in heavy-ion collisions using a thermal model), down to $\mu_\mathrm{B} = 0$, has been extensively studied, both theoretically \cite{Ding:2015ona,Borsanyi:2013bia} and experimentally \cite{Andronic:2017pug}. For vanishing and small $\mu_\mathrm{B}$, the transition from the low-temperature hadronic matter to the quark-gluon plasma at high temperature is well-known to be a smooth transition (crossover) \cite{Aoki:2006we}. At larger $\mu_\mathrm{B}$, however, it is generally expected that a first-order phase transition line exists, which ends in a second-order critical point belonging to the three-dimensional (3d) Ising model universality class \cite{Berges:1998rc,PhysRevD.58.096007,Asakawa:1989bq}. Comprehensive net-proton fluctuation studies have been carried out via the Beam Energy Scan (BES) program of the Relativistic Heavy Ion collider (RHIC) in the STAR experiment to search for the QCD critical point \cite{Pandav:2022xxx,Bzdak:2019pkr}. 

Recently, an ordering of the cumulant ratios of net-proton number,
$\frac{\chi_6}{\chi_2} < \frac{\chi_5}{\chi_1} < \frac{\chi_4}{\chi_2} < \frac{\chi_3}{\chi_1}$, has been reported by the STAR experiment across the energy range 7.7 to 200 GeV \cite{STAR:2022vlo} (corresponding to an estimated range of $\mu_\mathrm{B}\approx$ 20--450~MeV) where $\chi_n$ is the $n^\mathrm{th}$ order cumulant of event-by-event net-proton number distribution. This ordering was suggested previously by the Lattice-QCD (LQCD)
calculations for the cumulants of net-baryon number fluctuations at small values of $\mu_\mathrm{B}$ \cite{PhysRevD.101.074502,PhysRevD.96.074510}. Functional renormalization group (FRG) calculations were also found to follow this hierarchy of the inequalities \cite{PhysRevD.104.094047}. In contrast, predictions from the Hadron Resonance Gas (HRG) model, utilizing an ideal gas equation of state within a grand canonical ensemble framework, consistently yield positive unity for all cumulant ratios, lacking the observed ordering \cite{Borsanyi:2018grb}. One may wonder whether these inequalities are generically valid in the vicinity of any phase transition or even in wider regions surrounding it. In the QCD phase diagram, either near the cross over or near the QCD critical point, if it exists, whether such inequalities exist is an interesting question to ask. 

Relying on the anticipated universality of the QCD critical point with the 3d Ising model, we investigate in this study whether the observed ordering of higher-order cumulant ratios can arise in simpler two-dimensional (2d) spin models near continuous phase transitions. Quantitative properties of critical behavior, such as critical exponents and scaling functions, depend on dimensionality and are therefore not expected to match those of the QCD critical point. However, qualitative aspects of critical phenomena---such as divergence of the correlation length, enhanced fluctuations of the order parameter, and the scaling structure of higher-order cumulants---are common to systems undergoing second-order phase transitions. In particular, since higher-order susceptibilities correspond to derivatives of the free energy with respect to the external field, their qualitative behavior near criticality can often be explored in simpler statistical systems. The central question we address is therefore whether the ordering of cumulant ratios can emerge as a generic finite-size feature near criticality, or whether it relies on additional ingredients such as dimensionality, explicit symmetry breaking, or the presence of first-order transitions.

The expected QCD phase diagram, with a line of first order phase transitions at large $\mu_\mathrm{B}$ which ends in a critical point for smaller $\mu_\mathrm{B}$, motivates the use of simplified model system to probe the behavior of these inequalities in finite volumes expected in the heavy-ion collisions. The first order phase transition of 3d three-state Potts model in the absence of an external magnetic field \cite{PhysRevLett.43.799,Gavai:1988yq} weakens with the addition of a small magnetic field $(h)$ and leads to a critical point in the 3d Ising universality class at $h_c$ \cite{Karsch:2000xv}. Generically, increasing $h$ towards $h_c$ should be similar to decreasing $\mu_\mathrm{B}$ towards $\mu_{\mathrm{B},c}$ along the first order phase transition line, allowing one to learn more about the inequalities of the cumulant ratios. As a first step towards such an investigation, we have studied the higher-order cumulants of magnetization as a function of temperature for $h=0$ in the two-state ($\mathrm{Z}(2)$) and the three-state ($\mathrm{Z}(3)$) Potts model on 2d lattices in this paper. Both the models are known to have critical points albeit falling under different universality classes. The finite-size scaling of the higher-order cumulants of magnetization were investigated near the critical region in the respective models and the ordering of their ratios were tested. While these studies are ideally needed to be carried out in three dimensions in the presence of nonzero magnetic field, our work nevertheless provides a foundational analysis of the hierarchy of cumulant ratios, which can be generalized to the appropriate model. 

It is important to note that the cumulants studied in these spin models differ from those calculated in LQCD. In LQCD at $\mu_\mathrm{B} = 0$, baryon–antibaryon symmetry implies that all odd-order susceptibilities of the net-baryon number vanish exactly. Consequently, ratios involving odd cumulants become meaningful only when considering finite $\mu_\mathrm{B}$, typically through Taylor expansion coefficients of the pressure around $\mu_\mathrm{B}=0$~\cite{PhysRevD.101.074502}. In contrast, the symmetry structure of the Potts model does not enforce an exact analogue of baryon–antibaryon symmetry for the magnetization variable considered here, and therefore odd-order cumulants of magnetization need not vanish even at $h=0$. Comparison of this study to QCD should therefore be interpreted as qualitative.

The paper is structured as follows: in Sec.~\ref{sec:Models}, we provide an
overview of the $q$-state Potts models, detailing their formalism and the
computation of cumulants of magnetization up to $6^\mathrm{th}$ order in these models. We also discuss the application of finite size scaling relation, which enables the extraction of exponents associated with the scaling of higher-order cumulants. Section ~\ref{sec:NumericalSimulation} presents the
methodology employed for simulating the models and outlines the simulation
validation process. In Sec.~\ref{sec:Results}, we present the
temperature-dependent behavior of cumulants and their ratios. The exponents involved with the finite-size scaling of the higher-order cumulants are estimated in terms of known critical exponents of the model and compared with the numerical results. Finally, Sec.~\ref{sec:Conclusion} summarizes the key findings of our study.

\section{\label{sec:Models} Model and Observables}
The Hamiltonian for the the $q$-state Potts model \cite{Wu1982Potts} is given by
\begin{equation}
    \mathcal{H}=-\mathcal{J}\sum_{\langle i,j\rangle}\delta(\sigma_i,\sigma_j) - h\sum_{i}\delta(\sigma_i,0)\,,
\end{equation}
where $\mathcal{J}$ is a coupling constant in units of energy, determining the interaction strength among neighbouring spins, $h$ is the external magnetic field applied to the model and $\delta$ is the Kronecker delta function. Here, $\sigma_k$ can take any integer values in the set \{0, 1, 2, ..., ($q$ - 1)\} and for each $\sigma_k$, the spin orientations in the Potts model are represented by $s_k = \mathrm{exp}\left(\frac{i2\pi\sigma_k}{q}\right)$. The sum is over nearest
neighbors $\langle i,j\rangle$. In $d$-dimensions there are 2$d$ such pairs. We focus on
$d=2$ in this work. In the absence of an external magnetic field ($h$ = 0), the model exhibits an exact symmetry represented by the group of permutations $S_q$, which encompasses all possible global permutations of the $q$ spin values. This symmetry implies that each spin state is equally likely, and there is no preferred direction for alignment. However, below a critical temperature, the symmetry is spontaneously broken resulting in the alignment of spins along a specific direction and the establishment of an ordered phase. To measure the extent of symmetry breaking in the system, a quantity called the order parameter is used, that provides a measure of the overall alignment of spins in the model. The order parameter for the $q$-state Potts model ($m_\mathrm{Potts}$) that we use in this study is given by
\begin{equation}
    m_\mathrm{Potts} = \frac{q}{q-1}\left(\frac{\mathrm{max}(N_0, N_1, ..., N_{q-1})}{N} - \frac{1}{q} \right)\,,
\end{equation}
where $N_\alpha = \sum_{j=1}^{L^d}\delta(\sigma_i,\alpha)$ with $\alpha \in \{0, 1, 2, ..., q - 1\}$ represents the number of sites in the lattice with $\sigma_i = \alpha$ and $N = L^d = \sum_{\alpha}N_{\alpha}$ is the total number of spin sites in the lattice. 

In the following analysis and results, we will be dealing Potts model with both $q = 2$ and $q = 3$ states. The two-state Potts model is equivalent to the well-known Ising model for which the exact solution by Onsager gives the critical point at $T_C = \frac{2}{\mathrm{ln}(1+\sqrt{2})} \,\mathcal{J}/k_\mathrm{B} \approx 2.2692 \,\mathcal{J}/k_\mathrm{B}$ \cite{PhysRev.65.117}. The three-state Potts model that belongs to the same universality class as the 2d Ising model, undergoes a second order phase transition at $T_C = \frac{1}{\mathrm{ln}(1 + \sqrt{3})}$ $\mathcal{J}/k_\mathrm{B} \approx 0.9950 \,\mathcal{J}/k_\mathrm{B}$ \cite{Wu1982Potts}.

\subsection{\label{sec:Cumulants} Cumulants}
In this paper, the main objects of interest are the higher-order cumulants (also referred as susceptibilities in the paper), which reflect the higher-order correlations amongst the spins.
In spin models, the $p^\mathrm{th}$ order susceptibility normalized by the total number of spin sites in the lattice is defined \cite{pathria_statistical_2011} by
\begin{equation}
   \chi_p(T) = \frac{1}{L^d\beta^p}\left(\frac{\partial^p\mathrm{ln} Z}{\partial h^p}\right)_{h\to0}\,,
\end{equation}
where $\beta = 1/(k_\mathrm{B}T)$, $T$ is the temperature, $k_\mathrm{B}$ is the Boltzmann constant. Here, $Z$ is the partition function of the model given by its corresponding $\mathcal{H}$ as
\begin{equation}
    \label{eq:Z}
    Z=\sum_a\mathrm{exp}(-\beta\mathcal{H})\,.
\end{equation}
The sum over $a$ in Eq.~\ref{eq:Z} means the summation over all possible spin
configurations on the lattice. The higher-order spin-spin susceptibilities can be
associated with the cumulants of total magnetization ($M$). This is because $Z$ is proportional to the moment generating function \cite{PhysRevE.104.064103} of $M$ and therefore one can easily deduce, 
\begin{equation}
    \label{eq:susc_as_cumulants} \left(\frac{\partial^p\mathrm{ln} Z}{\partial h^p}\right)_{h\to0} = \beta^p K_p(M)\,.
\end{equation}

Here, $M$ is referred as magnetization which is equal to the product of the order parameter and the volume of the lattice, i.e., $M=L^d m_\mathrm{Potts}$. $K_p(M)$ is the $p^\mathrm{th}$ order cumulant of $M$. The expressions for $\chi_p$ (obtained by using the relation of cumulants in terms of central moments \cite{smith1995recursive}) up to $p=6$ is listed below,
\begin{eqnarray}
    \chi_1 &= \langle M\rangle/L^d \label{eq:chi1}\\ 
    \chi_2 &= \mu_2/L^d \label{eq:chi2} \\
    \chi_3 &= \mu_3/L^d \label{eq:chi3}\\
    \chi_4 &= (\mu_4-3\mu_2^2)/L^d \label{eq:chi4}\\
    \chi_5 &= (\mu_5-10\mu_3\mu_2)/L^d \label{eq:chi5}\\
    \chi_6 &= (\mu_6-15\mu_4\mu_2-10\mu_3^2+30\mu_2^3)/L^d  \label{eq:chi6}
\end{eqnarray}
where $\mu_p=\langle(M -\langle M\rangle)^p\rangle$ is the
$p^\mathrm{th}$ order central moment of the magnetization and the angular
brackets, $\langle ..\rangle$ represents average over all possible spin
configurations on the lattice. We will be studying these $\chi_p$'s as a
function of temperature in the critical region. %while also referring to them as higher order susceptibilities in the analysis as a generalization of the usual susceptibility of magnetization.

\subsection{\label{sec:FSS} Finite size scaling}

In this section, we obtain the generalized relation for finite size scaling of
the higher-order susceptibilities of magnetization in the Potts model. We
define, $t = (T-T_C)/T_C$ and $h_r = \beta h$. For a second order
(continuous) phase transition, there are at least two relevant scaling fields for
the free energy: temperature like and magnetic field like. The dependence of free energy on these scaling fields near the critical point can be established by a bulk Renormalization Group (RG) transformation. RG transformation involves length scale re-scaling, under which, the scaling fields generally look like $C_1 L^{1/\nu}t$ and $C_2 L^{\Delta/\nu} h_r$ respectively \cite{privman_fisher_1984,PATHRIA2011539}. Here, the exponent $\nu$ is defined such that $\xi
\sim t^{-\nu}$ (for a finite lattice $\xi \sim L$ at the critical point)
and $\Delta \equiv \beta + \gamma$ is such that $\chi_1 \sim t^\beta$ and $\chi_2\sim t^\gamma$ in the zero field limit, i.e., $h\rightarrow 0$. 

In the obtained scaling relation for the singular part of the free energy after RG transformations, irrelevant scaling fields and non-linear contributions from relevant scaling fields can be safely removed by setting them to zero. This simplification is valid under the condition that $d < d_{>}$, where $d_{>}$ is the upper critical dimension. Consequently, only the linear order of temperature-like ($t$) and magnetic field-like ($h_r$) scaling fields remains, serving as the sole parameters that determine the distribution of the system \cite{privman_fisher_1984}. The asymptotic finite size scaling relation of the singular part of \textit{reduced free energy density}, i.e $f_r^{(s)} = L^{-d} \ln Z$, can therefore be written as (for $t \rightarrow 0, h_r \rightarrow 0$)
\begin{equation} \label{eq:singular_free_energy_density}
    f_r^{(s)}(t, h; L) \approx L^{-d}Y(C_1 L^{1/\nu}t, C_2 L^{\Delta/\nu} h_r)\,.
\end{equation}
Here, the function $Y(x,y)$ is same for every system in the universality class, i.e. it is universal, while the coefficients $C_1$ and $C_2$ are generally not universal. The $n^{th}$ derivative of Eq.~\ref{eq:singular_free_energy_density} with respect to $h_r$ gives the scaling relation for the $n^{th}$ order finite-size susceptibility,
\begin{equation} \label{eq:ho_susceptibility_singular_part}
    \chi_n^{(s)}(t; L) \approx - C_2^n L^{\frac{n\Delta}{\nu} - d} \frac{\partial^n Y}{\partial h_r^n}\bigg\vert_{h_r = 0}\,.
\end{equation}
Using Eq.~(\ref{eq:ho_susceptibility_singular_part}), ratios of two susceptibilities can be expressed as
\begin{equation}
\frac{\chi_m}{\chi_n} =
C_2^{m-n} L^{\frac{(m-n)\Delta}{\nu}}
\frac{\partial^m Y/\partial h_r^m}{\partial^n Y/\partial h_r^n}
\bigg|_{h_r=0}.
\end{equation}
While the leading lattice-size dependence  is reduced in such ratios compared to individual susceptibilities, they retain contributions from the non-universal amplitude $C_2$ as well as from derivatives of the scaling function $Y$. Since scaling theory does not constrain the relative magnitudes or signs of the derivatives of $Y$, it does not imply any particular ordering among the cumulant ratios. Consequently, scaling theory by itself does not enforce any hierarchy among the cumulant ratios; their ordering depends on the detailed form of the scaling functions.

From Eq.~\ref{eq:ho_susceptibility_singular_part}, we can extract the scaling relation for the singular part of $\chi_n$, i.e., $\chi_n^{(s)} \sim L^{\frac{n\Delta}{\nu} - d}$. To simplify the exponents, we utilize the relations among the critical exponents: $\gamma = 2\Delta - (2-\alpha)$ and $2 - \alpha = d\nu$. Both of these relations are valid for $d=2$, but in general, the latter is only valid for any $d < d_>$. Thus, we can write the scaling exponent of $n^\mathrm{th}$ order susceptibility as,
\begin{equation} \label{eq:scaling exponents}
    e_n \equiv \frac{n\Delta}{\nu} - d = \frac{1}{2}\frac{n\gamma}{\nu} + (n-2)d\,.
\end{equation}
These exponents for $n$ = 2 to 6 can now be calculated using the respective values of $\gamma$ and $\nu$ for the model under consideration. For the $q=2$ state Potts model on a 2d lattice, we have $\gamma = \frac{7}{4}$ and $\nu = 1$ for $q=2$ \cite{Pelissetto:2000ek}, while $\gamma = \frac{13}{9}$ and $\nu = \frac{5}{6}$ for $q=3$ \cite{Wu1982Potts}.

\section{\label{sec:NumericalSimulation} Numerical simulation}
Both $q=2$ and $q=3$ state Potts models were simulated on 2d lattices using the Wolff Cluster algorithm in a region around the corresponding critical temperature ($T_C$) \cite{PhysRevLett.62.361}. All simulations were performed in the absence of an external magnetic field, i.e., $h$(or $h_r$) = 0. In the Wolff algorithm, clusters of similarly oriented spin sites are determined randomly which are then
flipped together to generate new states. This algorithm reduces the effect of critical slowing down near $T_C$ with increase in lattice size, i.e. it has a smaller dynamical exponent resulting in more efficient simulations in this region \cite{newman_barkema}. We have used periodic boundary conditions for the finite lattices, which are also commonly employed in simulations of LQCD at finite temperature and density \cite{DeGrand:2006zz}.
For $q=2$, square lattices of linear size $L$ $=$ 50, 60, 70, 80 and 90 were used in the simulations and during the simulation $10^4$ independent configurations
were generated after ensuring thermalization by throwing away sufficient initial
configurations. Similarly, for $q=3$, simulations were performed on square
lattices with $L$ $=$ 48, 64, 80, 96 and 128 to generate $2 \times 10^4$
independent configurations after appropriate thermalization.
Simulations were conducted for the temperature range $0.95T_C \le T \le 1.05T_C$, and the corresponding results are presented. Tables \ref{tab:TCL_Ising} and \ref{tab:TCL_Potts} show that this range adequately covers the estimated critical region for all lattice sizes examined. 
\begin{table}[htbp!]
    \centering
    %\captionsetup{font=scriptsize}
    
    \begin{tabular}{@{}c@{\hspace{0.15in}}c@{\hspace{0.15in}}c@{\hspace{0.15in}}c@{}}\hline
        Lattice size &  $T_{C,L}$ &  $\chi_2(T_{C,L})$ & $\sigma_{C,L}$ \\ \hline
        50 & 2.3151 $\pm$ 0.0007 & 105.656 & 0.0495 $\pm$ 0.0011 \\
        60 & 2.3076 $\pm$ 0.0006 & 143.913 & 0.0424 $\pm$ 0.0009 \\
        70 & 2.3020 $\pm$ 0.0007 & 188.497 & 0.0360 $\pm$ 0.0009 \\
        80 & 2.2974 $\pm$ 0.0005 & 238.766 & 0.0310 $\pm$ 0.0007 \\
        90 & 2.2956 $\pm$ 0.0005 & 289.739 & 0.0288 $\pm$ 0.0007 \\ \hline
    \end{tabular}
    \caption{The extracted values for position ($T_{C,L}$), height ($\chi_2(T_{C,L})$) and width ($\sigma_{C,L}$) of the peak of $\chi_2$ distribution on an $L^{2}$ lattice as a function of temperature for $q = 2$ state Potts model. The quoted values after '$\pm$' represent statistical uncertainties derived from the fit.  }
    \label{tab:TCL_Ising}
\end{table}

\begin{table}[htbp!]
    \centering
    %\captionsetup{font=scriptsize}
    
    \begin{tabular}{@{}c@{\hspace{0.15in}}c@{\hspace{0.15in}}c@{\hspace{0.15in}}c@{}}\hline
        Lattice size &  $T_{C,L}$ &  $\chi_2(T_{C,L})$ & $\sigma_{C,L}$ \\ \hline
        48 & 1.0039 $\pm$ 0.0001 & 105.656 & 0.0102 $\pm$ 0.0002 \\
        64 & 1.0012 $\pm$ 0.0001 & 143.913 & 0.0071 $\pm$ 0.0002 \\
        80 & 0.9998 $\pm$ 0.0001 & 188.497 & 0.0056 $\pm$ 0.0001 \\
        96 & 0.9989 $\pm$ 0.0001 & 238.766 & 0.0046 $\pm$ 0.0001 \\
        128 & 0.9978 $\pm$ 0.0001 & 289.739 & 0.0033 $\pm$ 0.0001 \\ \hline
    \end{tabular}
    \caption{The extracted values for position ($T_{C,L}$), height ($\chi_2(T_{C,L})$) and width ($\sigma_{C,L}$) of the peak of $\chi_2$ distribution on an $L^{2}$ lattice as a function of temperature for $q = 3$ state Potts model. The quoted values after '$\pm$' represent statistical uncertainties derived from the fit. }
    \label{tab:TCL_Potts}
\end{table}
In order to obtain independent spin configurations on the lattice, configurations were recorded at the interval of $2\tau$ where $\tau$ is the auto-correlation time that is obtained from the auto-correlation function (of $M$) defined in Eq.~\ref{eq:auto_corr} using an exponential decay ansatz $\phi(t) \sim \mathrm{exp}\left({-\frac{t}{\tau}}\right)$.
\begin{equation} \label{eq:auto_corr}
    \phi(t)=\frac{\langle M(0)M(t)\rangle - \langle M\rangle^2}{\langle M^2\rangle - \langle M\rangle^2}
\end{equation}
The magnetization is calculated for each spin configuration in the simulation and the higher-order susceptibilities of magnetization are calculated using Eqs.~\ref{eq:chi1} - \ref{eq:chi6} by averaging over all possible spin configurations on the lattice.

\section{\label{sec:Results} Results}
\begin{figure}[htbp!]
  \centering
  \resizebox{0.45\linewidth}{!}{%
    \includegraphics{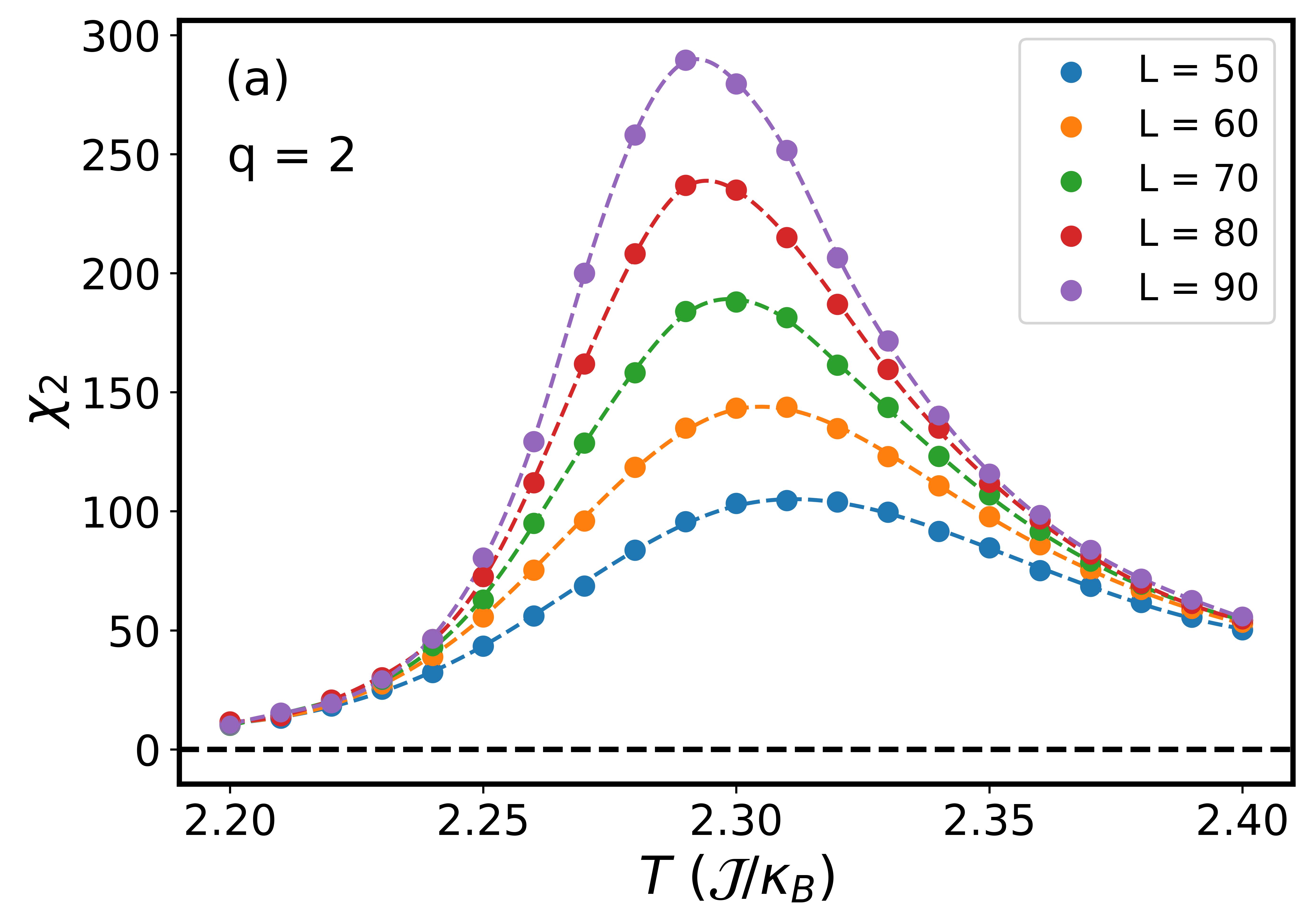}
  }
  \resizebox{0.49\linewidth}{!}{%
    \includegraphics{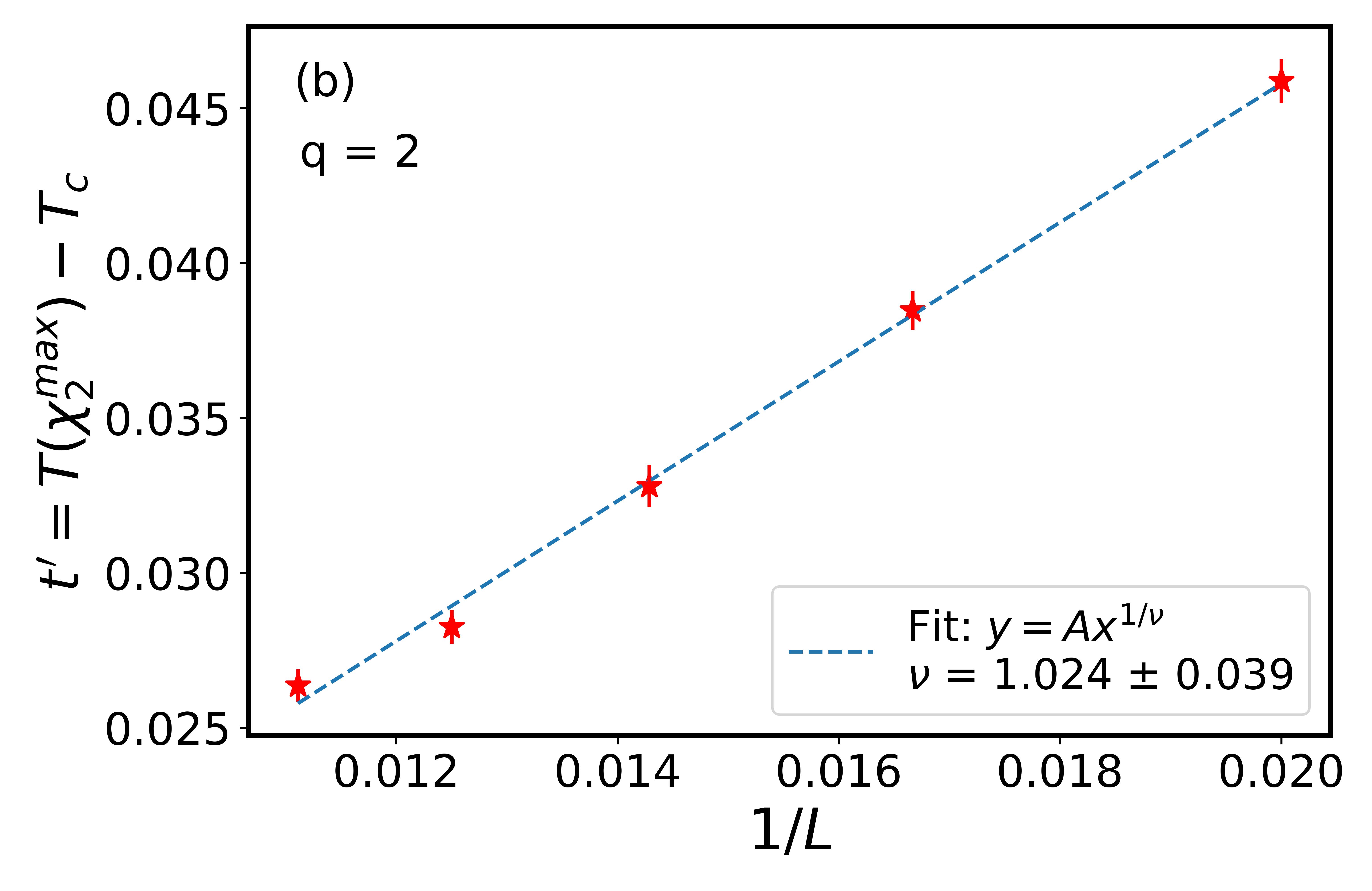}
  }
  \caption{(a) The second order susceptibility of magnetization for the 2d Ising model (equivalent to $q = 2$ state Potts model) shown as a function of temperature. The results shown by different colored markers represent different lattice sizes ($L$). The dotted lines connecting the points are intended to guide the viewer's eye. (b) The reduced temperature, $t'=T(\chi_2^\mathrm{max})-T_{C}$ is plotted as a function of $1/L$, and fitted with the function $y=Ax^{1/\nu}$ to extract the value of $\nu$.}
  \label{fig:plots_chi2_and_fitIsing}
\end{figure}

\begin{figure}[htbp!]
  \centering
  \resizebox{0.45\linewidth}{!}{%
    \includegraphics[width=\linewidth]{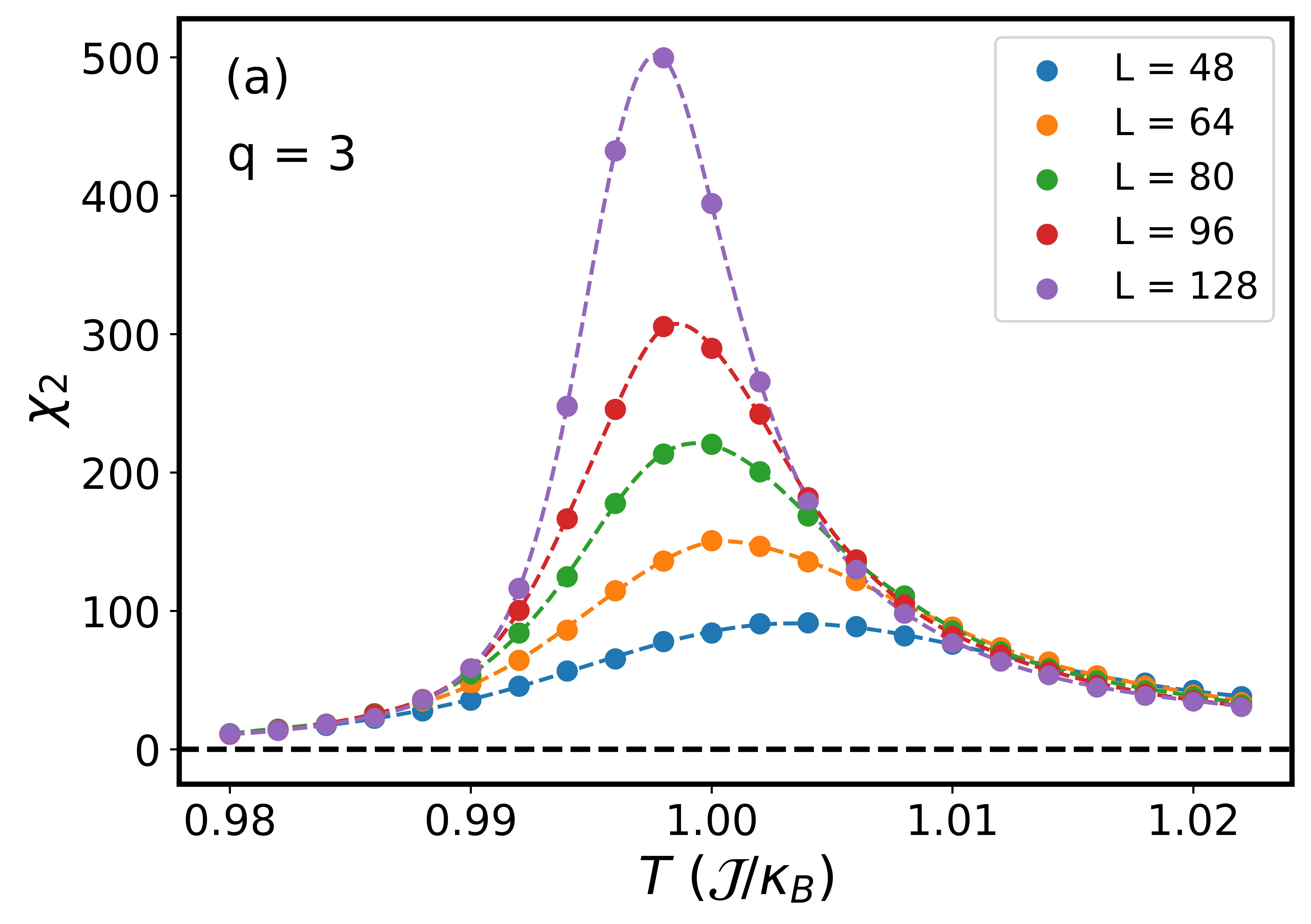}
  }
  \resizebox{0.49\linewidth}{!}{%
    \includegraphics[width=\linewidth]{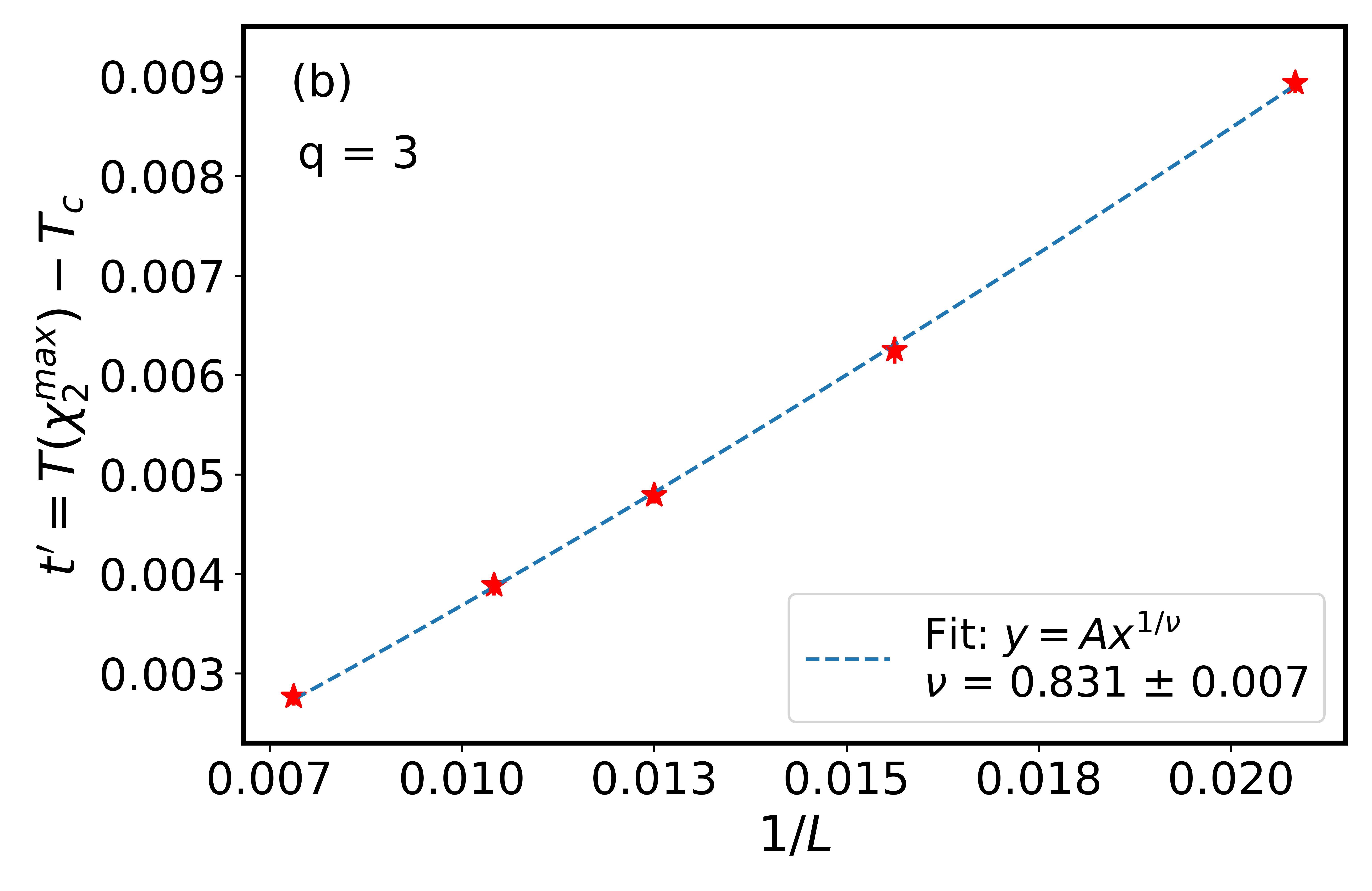}
  }
  \caption{(a) The second order susceptibility of magnetization for the 2d $q = 3$ state Potts model shown as a function of temperature. The results shown by different colored markers represent different lattice sizes ($L$). The dotted lines connecting the points are intended to guide the viewer's eye. (b) The reduced temperature, $t'=T(C_2^\mathrm{max})-T_{C}$ is plotted as a function of $1/L$, and fitted with the function $y=Ax^{1/\nu}$ to extract the value of $\nu$.}
    \label{fig:plots_chi2_and_fitPotts}
\end{figure}

The top panels of Fig.~\ref{fig:plots_chi2_and_fitIsing} and 
Fig.~\ref{fig:plots_chi2_and_fitPotts} display 
the second-order susceptibility, denoted as $\chi_2$, for the 
$q = 2$ and $q=3$ state Potts model respectively with varying lattice sizes, $L$. In both cases, $\chi_2$ exhibits a distinctive peak structure. As may be expected, the locations ($T_{C,L}$) and 
magnitudes ($\chi_2(T_{C,L})$) of these peaks vary with changes in the lattice size. Their scaling behavior can be utilized to extract the critical exponents, specifically $\gamma$ and $\nu$, which will be further elaborated upon in the subsequent section, \ref{sec:CriticalExponents}.

\subsection{\label{sec:CriticalExponents} Critical exponents}
The critical exponents, $\gamma$ and $\nu$ can be obtained using the scaling
relations, $\chi_2(T_{C,L}) \propto L^{-\gamma/\nu}$ and $T_{C,L} - T_{C}
\propto L^{-1/\nu}$, where $T_{C,L}$ is the pseudo-critical temperature for a 
given lattice size, and $T_C$ represents the critical temperature of the
model in the limit $ L \to \infty$. The pseudo-critical temperature is an effective critical temperature observed in finite-sized systems, reflecting behavior similar to the critical point in infinite-sized systems. $T_{C,L}$ is defined as the location of the peak in the second order
susceptibility, $\chi_2$ distribution as shown in
Figs.~\ref{fig:plots_chi2_and_fitIsing} and ~\ref{fig:plots_chi2_and_fitPotts}.
By fitting $\chi_2$ as a Gaussian function of the temperature, $T$, 
the peak location ($T_{C,L}$, $\chi_2(T_{C,L})$)
is determined for each lattice size. The obtained values of
$T_{C,L}$ and $\chi_2(T_{C,L})$ as a function of $L$ are tabulated in Tables \ref{tab:TCL_Ising} and \ref{tab:TCL_Potts} for $q=$ 2 and $q=$ 3 state Potts model respectively. Also, we define the critical region around $T_{C,L}$ for each lattice size by the width of the $\chi_2$ distribution peak, which is equal to the $\sigma$ of the fitted Gaussian distribution function, represented here as $\sigma_{C,L}$. Using the values of $T_{C,L}$, the difference $t'= T_{C,L} - T_{C}$ is plotted as a function of $1/L$, and fitted with its scaling relations for both $q=2$ and $q=3$ state Potts model as shown in the bottom panels of Figs.~\ref{fig:plots_chi2_and_fitIsing} and \ref{fig:plots_chi2_and_fitPotts} respectively. Alternatively, $T_C$ can be estimated by plotting directly $T_{C,L}$ against $L$ and fitting with $y=A+Bx^{-1/\nu}$. The resulting estimates ($T_C = 2.2780\pm0.0089$ for $q=2$ and $T_C = 0.9953\pm0.0001$ for $q=3$) are found to be consistent with the known values \cite{PhysRev.65.117,Wu1982Potts} within uncertainties. For precise measurements of the critical exponents, however we employ the known values of $T_{C}$ in the former method and extracted the coefficient $\nu$. The peak position, $\chi_2(T_{C,L})$ as a function of $L$, is also fitted with its scaling relation. Therefore, the extracted values of $\gamma$ and $\nu$ from the fits with the scaling relations are as follows: for $q=2$, $\gamma=1.758\pm0.060$, $\nu=1.019\pm0.034$ whereas for $q=3$, $\gamma=1.435\pm0.014$, $\nu=0.830\pm0.008$. The excellent agreement observed between the calculated values of $\gamma$ and $\nu$ and their theoretical counterparts affirms the validation of our simulation.

\begin{figure*}[htbp!]
    \centering
    \includegraphics[width=0.46\textwidth]{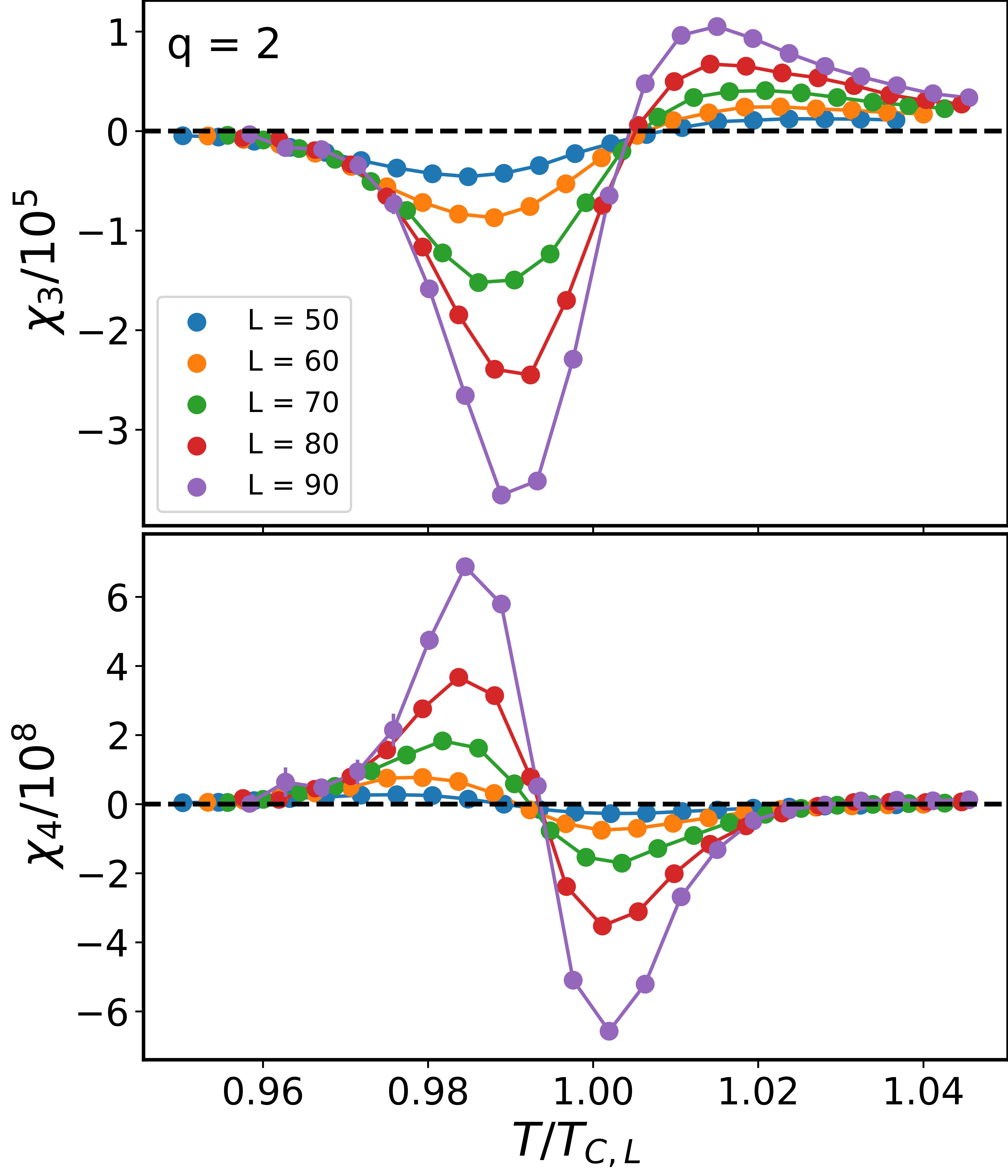}
    \hspace{0.02\textwidth}
    \includegraphics[width=0.46\textwidth]{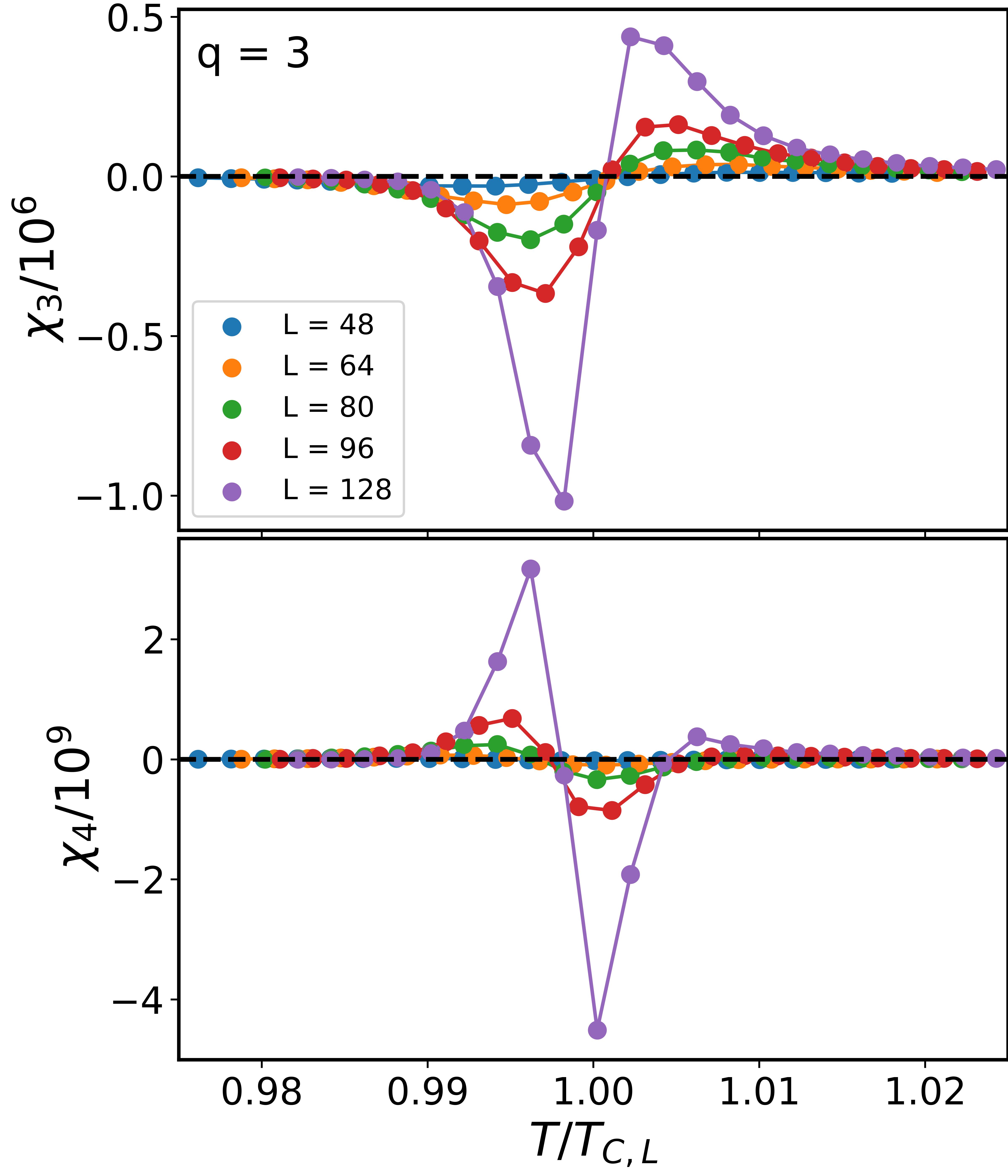}
    
    \caption{The third and fourth order susceptibilities of magnetization, $\chi_3$ and $\chi_4$, are shown for different lattice sizes as a function of temperature scaled by critical temperature of the corresponding lattice, $T/T_{C,L}$. The left side plot belongs to 2d $q=2$ state Potts model whereas the right side plot is for $q=3$ state Potts model. The upper panel represents $\chi_3$ and the bottom panel represents $\chi_4$ [where $\chi_3$ and $\chi_4$ divided by a factor $10^5$ ($10^6$) and $10^8$ ($10^9$) respectively for $q=2$ ($q=3$)]. }
    \label{fig:plots_susc_order_3&4}
\end{figure*}

\begin{figure*}[htbp!]
    \centering
    \includegraphics[width=0.46\textwidth]{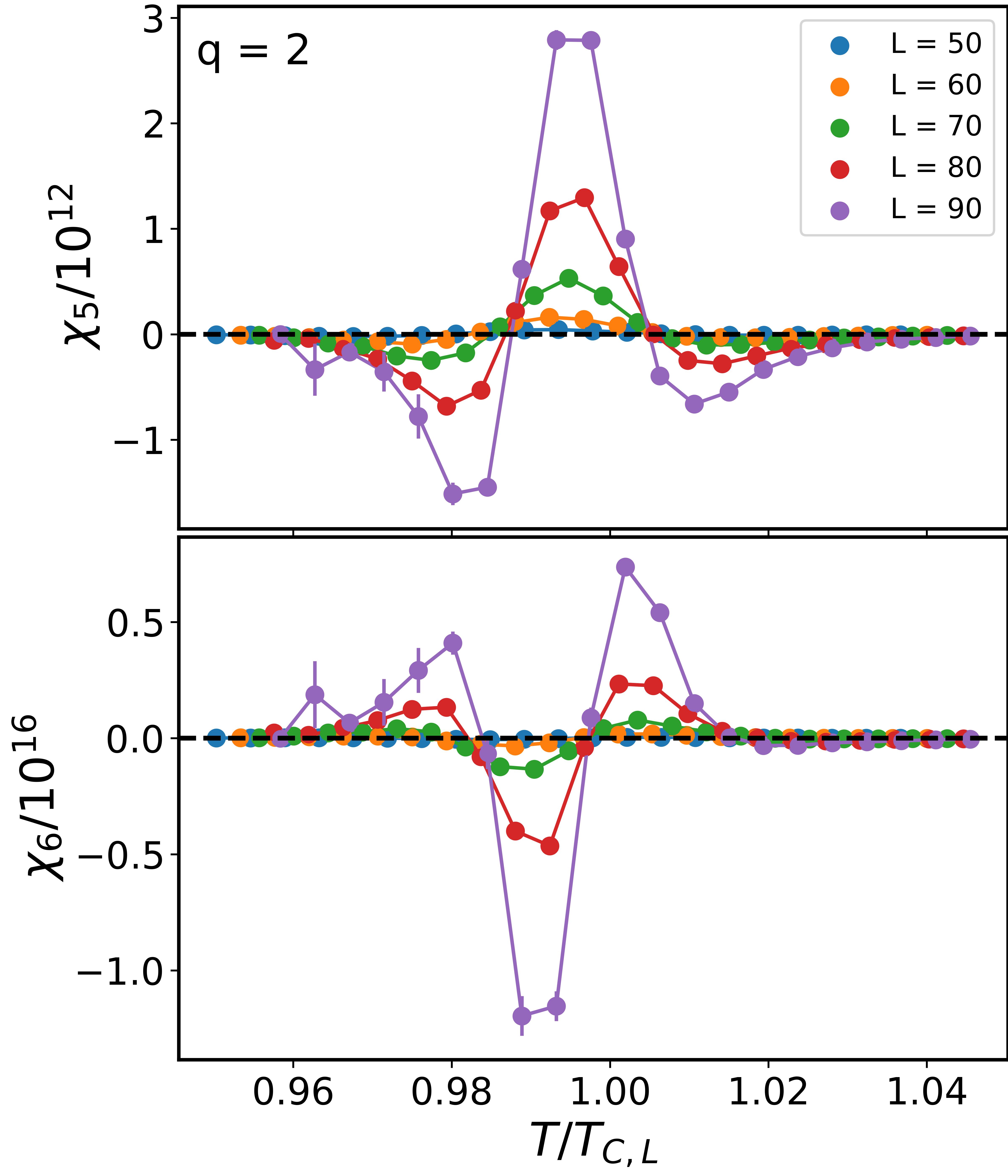}
    \hspace{0.02\textwidth}
    \includegraphics[width=0.46\textwidth]{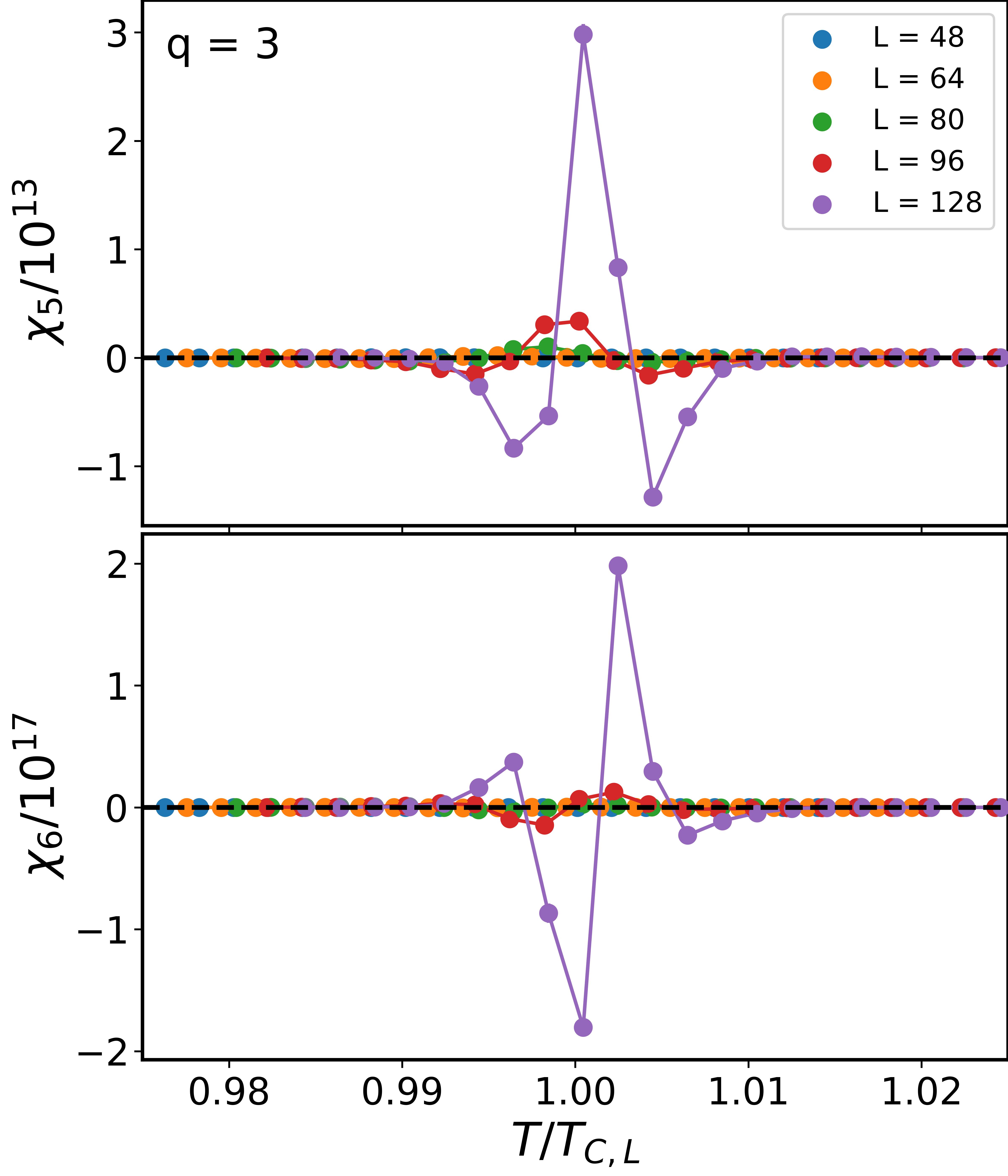}
    
    \caption{The fifth and sixth order susceptibilities of magnetization, $\chi_5$ and $\chi_6$, are shown for different lattice sizes as a function of temperature scaled by critical temperature of the corresponding lattice, $T/T_{C,L}$. The left side plot belongs to 2d $q=2$ state Potts model whereas the right side plot is for $q=3$ state Potts model. The upper panel represents $\chi_5$ and the bottom panel represents $\chi_6$ [where $\chi_5$ and $\chi_6$ divided by a factor $10^{12}$ ($10^{13}$) and $10^{16}$ ($10^{17}$) respectively for $q=2$ ($q=3$)]. }
    \label{fig:plots_susc_order_5&6}
\end{figure*}

We plot the higher-order susceptibilities, $\chi_3$ and $\chi_4$ for
both $q=2$ and $q=3$ state Potts model in Fig.~\ref{fig:plots_susc_order_3&4}
as a function of temperature scaled by $T_{C,L}$. Since, these $\chi_n$'s are
one order derivative of its preceding $\chi_{n-1}$ (for $n > 2$), the sign of
peak structure gets reversed in $\chi_n$ with respect to $\chi_{n-1}$. Similarly, $\chi_5$ and $\chi_6$ are shown as a function of $T/T_{C,L}$ for both the models in Fig.~\ref{fig:plots_susc_order_5&6}. The
height of the peaks or dips in these higher-order susceptibilities are found to be lattice size dependent. They also follow a scaling as a function of $L$ as discussed in
section \ref{sec:FSS}. In order to determine the scaling exponents of the higher-order susceptibilities, we fit the maximum or minimum value of $\chi_n$, depending on which is prominent, as a
function of $L$ against the scaling relation: $\chi_{n}^\mathrm{max/min} \propto L^{b}$. The scaling exponents of $\chi_3$, $\chi_4$, $\chi_5$ and $\chi_6$
obtained from the fit are referred as $e_n^{''}$. For $q=2$ state Potts model, the values of scaling exponents are given in
Table~\ref{tab:higher_order_susc_exponents_Ising}, whereas for $q=3$ state Potts
model, they are provided in Table~\ref{tab:higher_order_susc_exponents_potts}. Additionally, we
compare the scaling exponents (Eq.~\ref{eq:scaling exponents}) obtained using
exact values of $\gamma$ and $\nu$ as well as by using the estimates of $\gamma$
and $\nu$ (from scaling of $\chi_2$) obtained in this paper, and they are
referred as $e_n$ and $e_n'$ respectively in
Tables~\ref{tab:higher_order_susc_exponents_Ising} and
\ref{tab:higher_order_susc_exponents_potts}. The numerically estimated scaling
exponents, $e_n''$ for the higher-order
susceptibilities show a good agreement with the analytically obtained values. We
also note that determining the exponents from finite size scaling of $\chi_n$'s instead of
using the numerical estimate of $\gamma$ and $\nu$ from $\chi_2$ result in
smaller numerical errors.

\begin{table}[htbp!]
    \centering
    %\captionsetup{font=scriptsize}
    
    \begin{tabular}{@{}c@{\hspace{0.2in}}c@{\hspace{0.2in}}c@{\hspace{0.2in}}c@{}}\hline
        Quantity & $e_n$ & $e_n'$ & $e_n''$ \\ \hline
%        2 & $\chi_2 \equiv \chi_2$ & 1.75 &  $1.757 \pm 0.035$\\
        $\chi_3$ & 3.625 & $3.566 \pm 0.097$ & $3.608 \pm 0.026$ \\
        $\chi_4$ & 5.500 & $5.420 \pm 0.130$ & $5.500 \pm 0.074$ \\
        $\chi_5$ & 7.375 & $7.280 \pm 0.160$ & $7.257 \pm 0.067$ \\
        $\chi_6$ & 9.250 & $9.130 \pm 0.190$ & $9.196 \pm 0.066$ \\ \hline
    \end{tabular}
    \caption{Finite size scaling exponents of higher-order susceptibilities, $e_n^{''}$ obtained from fits to peak (or dip) values of $\chi_n$ as a function of lattice size in 2d $q = 2$ state Potts model. Here, $e_n$ is computed based on exact values of $\gamma$ and $\nu$ ($\gamma = 7/4$ and $\nu = 1$) in Eq.~\ref{eq:scaling exponents}, while $e_n'$ is derived using our estimates of $\gamma$ and $\nu$ discussed in Sec.~\ref{sec:CriticalExponents}. The quoted values after '$\pm$' represent statistical uncertainties derived from the fit. }
    \label{tab:higher_order_susc_exponents_Ising}
\end{table}

\begin{table}[htbp!]
    \centering
    %\captionsetup{font=scriptsize}
    
    \begin{tabular}{@{}c@{\hspace{0.2in}}c@{\hspace{0.2in}}c@{\hspace{0.2in}}c@{}}\hline
        Quantity & $e_n$ & $e_n'$ & $e_n''$ \\ \hline
%        2 & $\chi_2 \equiv \chi_2$ & 1.733 &  $1.732 \pm 0.005$\\
        $\chi_3$ & 3.599 & $3.593 \pm 0.036$ & $3.619 \pm 0.021$ \\
        $\chi_4$ & 5.466 & $5.458 \pm 0.047$ & $5.501 \pm 0.047$ \\
        $\chi_5$ & 7.333 & $7.322 \pm 0.059$ & $7.301 \pm 0.047$ \\
        $\chi_6$ & 9.199 & $9.187 \pm 0.071$ & $9.197 \pm 0.077$ \\ \hline
    \end{tabular}
    \caption{Finite size scaling exponents of higher-order susceptibilities, $e_n^{''}$ obtained from fits to peak (or dip) values of $\chi_n$ as a function of lattice size in 2d $q = 3$ state Potts model. Here, $e_n$ is computed based on exact values of $\gamma$ and $\nu$ ($\gamma = 13/9$ and $\nu = 5/6$) in Eq.~\ref{eq:scaling exponents}, while $e_n'$ is derived using our estimates of $\gamma$ and $\nu$ discussed in Sec.~\ref{sec:CriticalExponents}. The quoted values after '$\pm$' represent statistical uncertainties derived from the fit. }
    \label{tab:higher_order_susc_exponents_potts}
\end{table}

\subsection{\label{sec:RatioCumulants} Ordering of Cumulant ratios}
\begin{figure}
    \centering
    \includegraphics[width=0.42\linewidth]{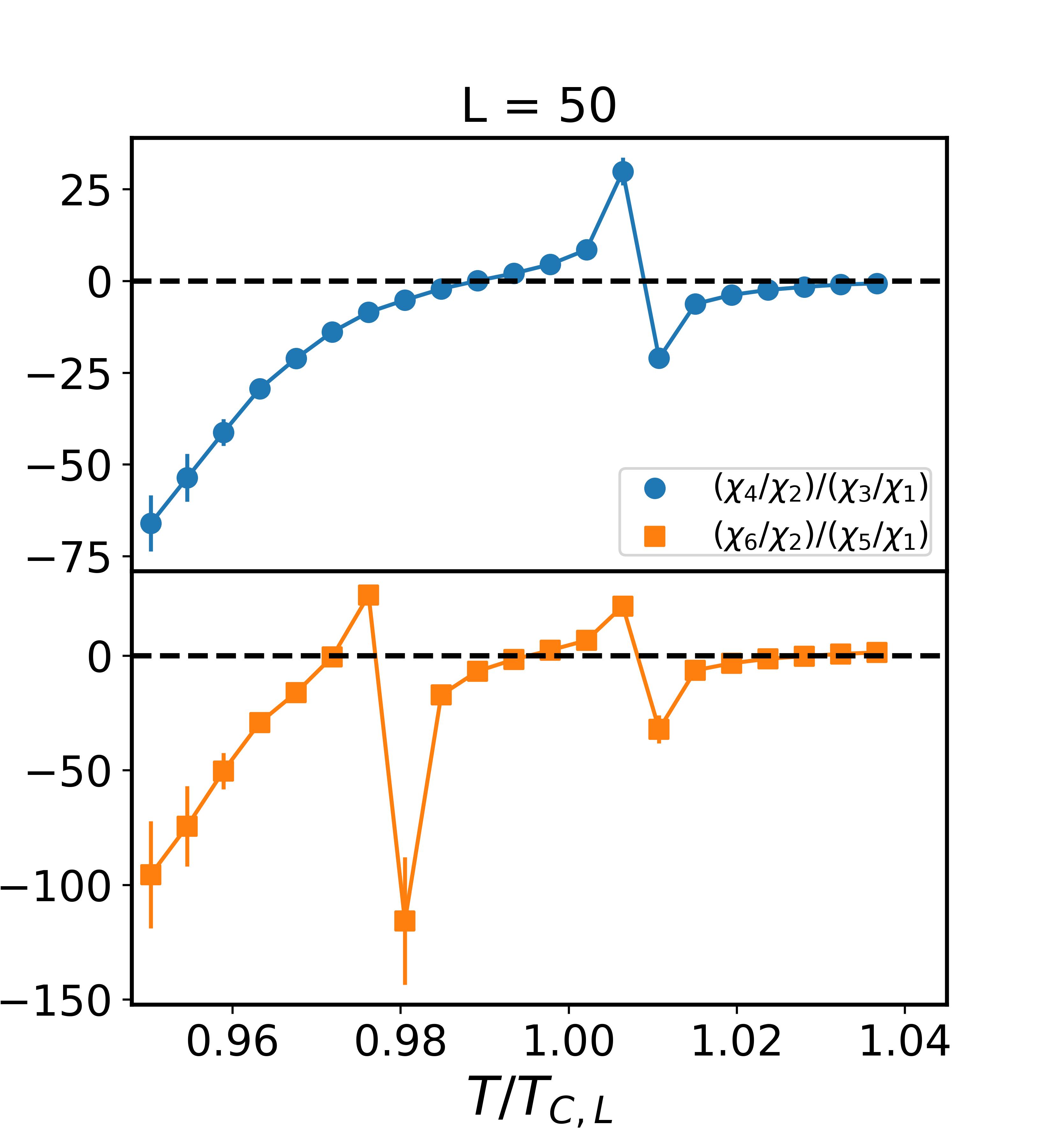}
    \caption{Double ratios $(\chi_6/\chi_2)/(\chi_5/\chi_1)$ and $(\chi_4/\chi_2)/(\chi_3/\chi_1)$ as a function of temperature scaled by critical temperature of the corresponding lattice, $T/T_{C,L}$ in $q = 2$ state Potts model. }
    \label{fig:doubleratiosIsing}
\end{figure}

\begin{figure}
    \centering
    \includegraphics[width=0.42\linewidth]{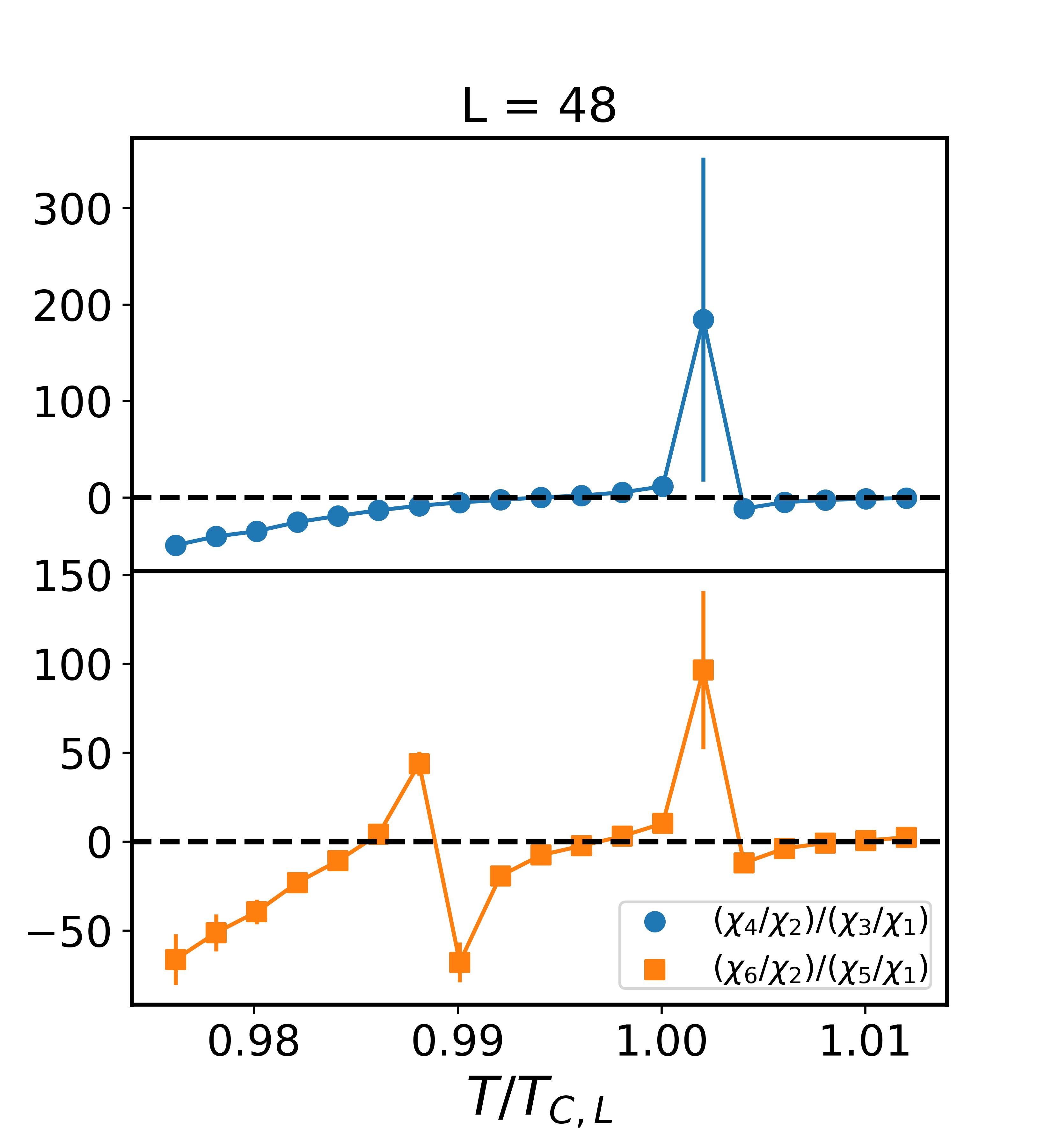}
    \caption{Double ratios $(\chi_6/\chi_2)/(\chi_5/\chi_1)$ and $(\chi_4/\chi_2)/(\chi_3/\chi_1)$ as a function of temperature scaled by critical temperature of the corresponding lattice, $T/T_{C,L}$ in $q = 3$ state Potts model. }
    \label{fig:doubleratiosPottsq3}
\end{figure}

%% Cumulant Ratios %%%%%%%%%%%%
In this section, we delve into the discussion of whether the 2d $q=2$ and $q=3$ state Potts models exhibit ordering in the susceptibility ratios, $\frac{\chi_6}{\chi_2}<\frac{\chi_5}{\chi_1}<\frac{\chi_4}{\chi_2}<\frac{\chi_3}{\chi_1}$. While it may apparently seem reasonable to only compare susceptibility ratios that are proportional to the same powers of $L$ (such as $\frac{\chi_6}{\chi_2}$ and $\frac{\chi_5}{\chi_1}$ proportional to $L^{4}$, or $\frac{\chi_4}{\chi_2}$ and $\frac{\chi_3}{\chi_1}$ proportional to $L^{2}$), we cannot overlook the terms $C_{2}^{n}$ and $\frac{\partial^nY}{\partial h_r^n}\bigg\vert_{h_r=0}$ in Eq.~\ref{eq:ho_susceptibility_singular_part}. In Fig.\ref{fig:doubleratiosIsing}, we plot the double ratio of susceptibilities, $\frac{\chi_4}{\chi_2}/\frac{\chi_3}{\chi_1}$ and $\frac{\chi_6}{\chi_2}/\frac{\chi_5}{\chi_1}$, for the $q=2$ state Potts model as a function of temperature scaled by $T_{C,L}$ in the top and bottom panels respectively. Similar results for the $q=3$ state Potts model can be seen in Fig.\ref{fig:doubleratiosPottsq3}. In both the $q=2$ and $q=3$ state Potts models, $\chi_6/\chi_2$ ($\chi_4/\chi_2$) is not always less than $\chi_5/\chi_1$ ($\chi_3/\chi_1$) across the studied temperature range. Despite the cancellation of $L$ dependence in the double ratios, the influence of $C_{2}^{n}$ and $\frac{\partial^nY}{\partial h_r^n}\bigg\vert_{h_r=0}$ is significant in determining the inequality among the ratios. Therefore, it is crucial to have a comprehensive comparison of the entire set of inequalities in these spin models.

\begin{figure*}[htbp!]
  \centering
    \includegraphics[width=0.325\linewidth]{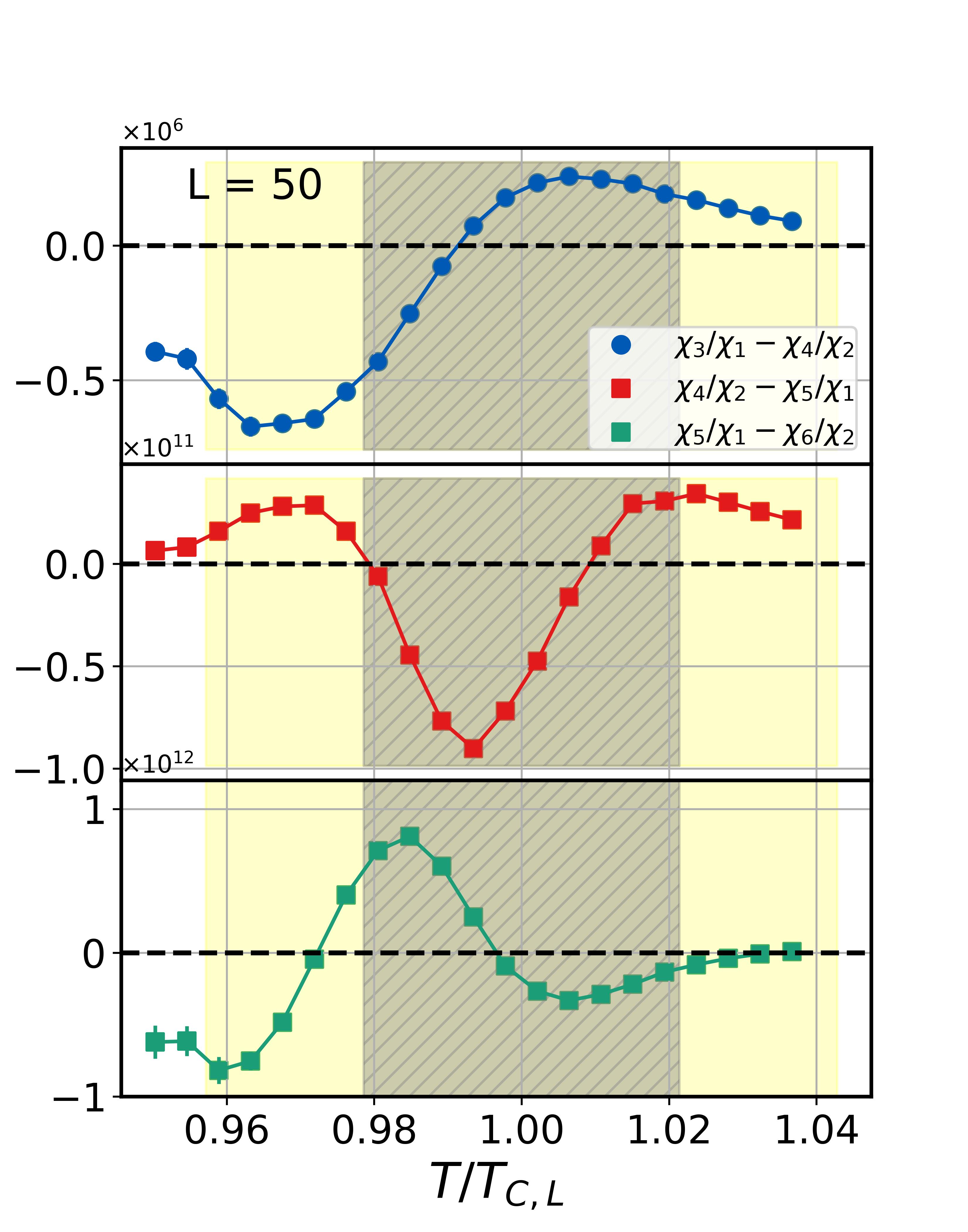}
    \includegraphics[width=0.325\linewidth]{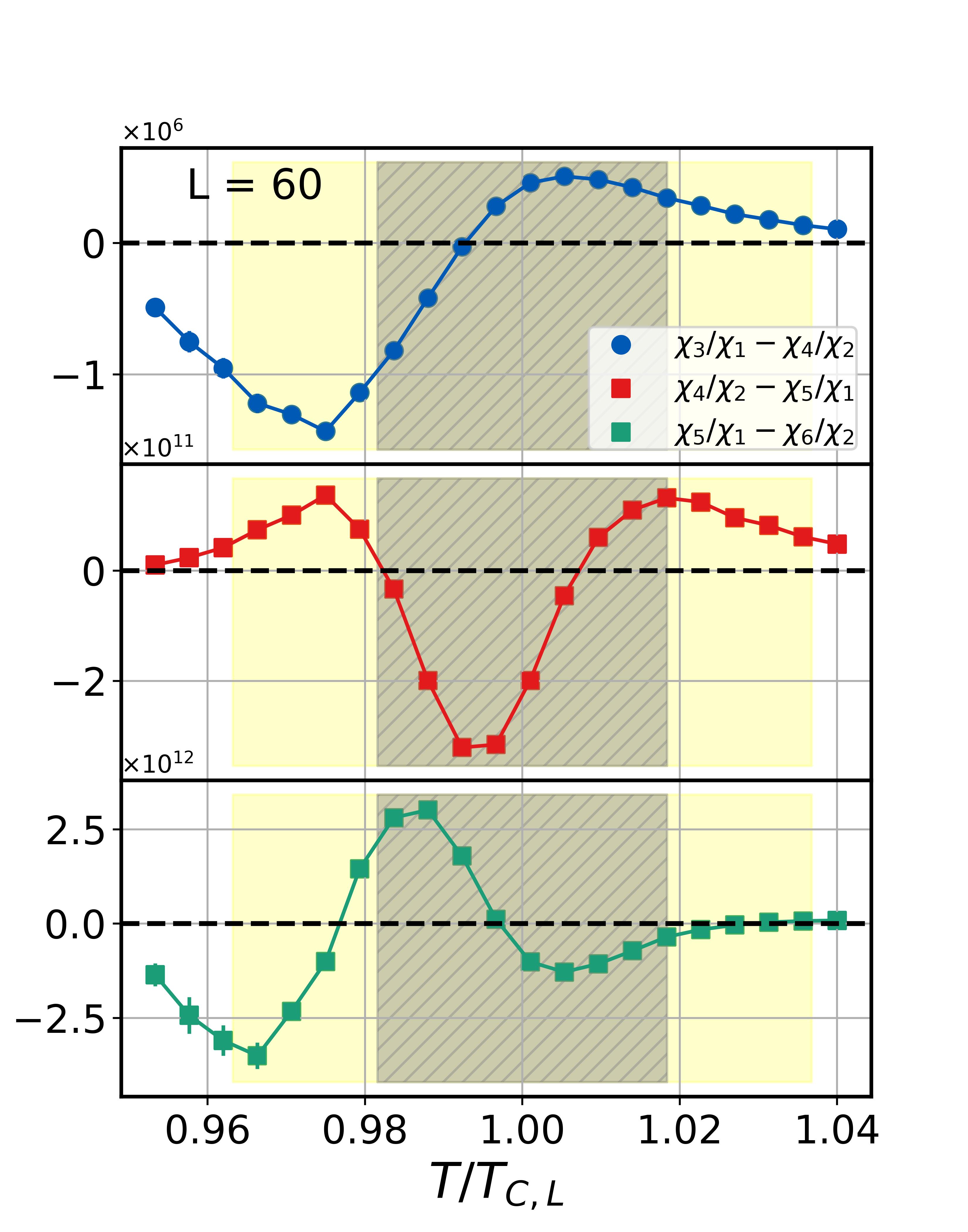}
    \includegraphics[width=0.325\linewidth]{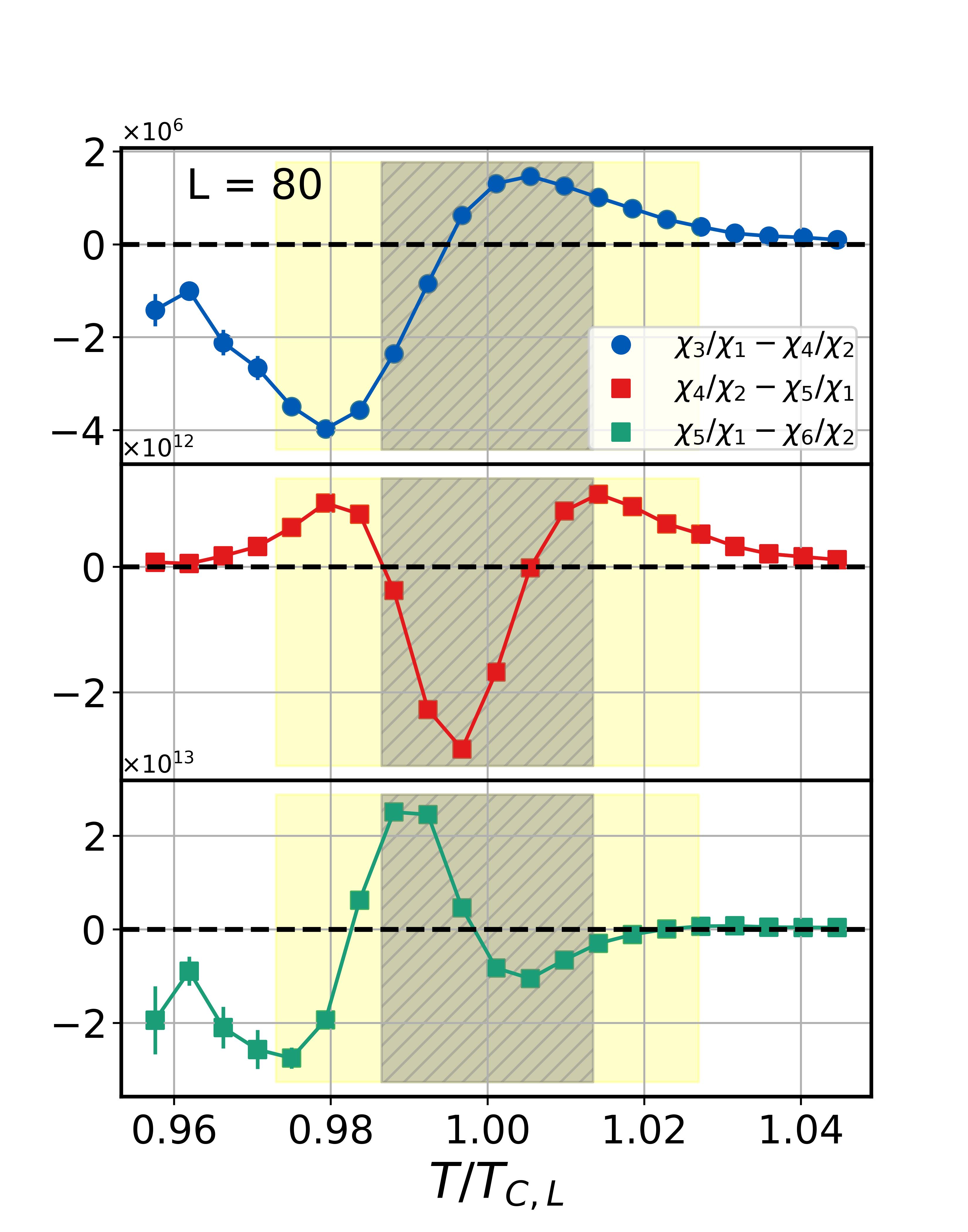} 
    \caption{The difference among ratios of the susceptibilities, ($\chi_3/\chi_1-\chi_4/\chi_2$), ($\chi_4/\chi_2-\chi_5/\chi_1$) and ($\chi_5/\chi_1-\chi_6/\chi_2$) for 2d $q = 2$ state Potts model with lattice sizes, L = 50 (left plot), 60 (middle plot), and 80 (right plot) are shown as a function of scaled temperature, $T/T_{C,L}$. The yellow shaded band indicates the critical region defined by $T_{C,L} \pm 2\sigma_{C,L}$, while the grey hatched region highlights the narrower interval $T_{C,L} \pm \sigma_{C,L}$.}
    \label{fig:Ising_ratios}
\end{figure*}

Figure~\ref{fig:Ising_ratios} presents the differences between successive susceptibility ratios, $\frac{\chi_3}{\chi_1}-\frac{\chi_4}{\chi_2}$, $\frac{\chi_4}{\chi_2}-\frac{\chi_5}{\chi_1}$, and $\frac{\chi_5}{\chi_1}-\frac{\chi_6}{\chi_2}$, in the $q=2$ state Potts model for lattice sizes of 50, 60, and 80 as a function of $T/T_{C,L}$. For completeness, the individual ratios, $\frac{\chi_3}{\chi_1}$, $\frac{\chi_4}{\chi_2}$, $\frac{\chi_5}{\chi_1}$, and $\frac{\chi_6}{\chi_2}$ are shown in the Appendix (Fig.~\ref{fig:Ising_ratios_app}). The yellow and grey shaded area indicate the 2$\sigma$ and 1$\sigma$ bands of the critical region, defined by $T_{C,L}\pm2\sigma_{C,L}$ and $T_{C,L}\pm\sigma_{C,L}$, respectively, with $\sigma_{C,L}$ values listed in Table \ref{tab:TCL_Ising}. As seen in Figs.~\ref{fig:plots_susc_order_3&4} and \ref{fig:plots_susc_order_5&6}, the magnitudes of higher-order susceptibilities differ by more than $\sim\mathcal{O}(10^3)$, making direct comparison of the ratios less transparent. To assess their relative ordering, we instead examine the differences between successive ratios (in the sequence $\frac{\chi_3}{\chi_1}$, $\frac{\chi_4}{\chi_2}$, $\frac{\chi_5}{\chi_1}$, $\frac{\chi_6}{\chi_2}$), where the sign directly reflects the hierarchy. Positive values of all differences in a given temperature interval imply that the ordering is satisfied. The statistical uncertainties of the cumulants and their ratios are estimated using a bootstrap resampling procedure with $\sim100$ samples, applied to statistically independent sample obtained by recording configurations at intervals of $2\tau$ (see Sec.~\ref{sec:NumericalSimulation}). To quantify the statistical significance of the inequalities among cumulant ratios, we examine the distributions of the differences between the relevant ratios (e.g., $\chi_3/\chi_1 - \chi_4/\chi_2$ for the inequality $\chi_3/\chi_1<\chi_4/\chi_2$). The robustness of a given inequality is assessed based on whether the sign of the corresponding difference is preserved within the estimated statistical uncertainties.

These differences exhibit clear sign changes in the vicinity of $T_{C,L}$. A clear ordering, $\frac{\chi_5}{\chi_1} < \frac{\chi_4}{\chi_2} < \frac{\chi_3}{\chi_1}$, is observed on the high-temperature side ($T > T_{C,L}$), as indicated by the positive values of the corresponding differences in the top and middle panels of Fig.~\ref{fig:Ising_ratios}, consistently across all lattice sizes. This ordering sets in at $T \gtrsim T_a$, with $T_a \approx 1.01T_{C,L}$, and is observed within both the $1\sigma$ and $2\sigma$ regions. In contrast, the expected hierarchy $\frac{\chi_6}{\chi_2} < \frac{\chi_5}{\chi_1}$ is not observed across most of the critical region, where $\frac{\chi_5}{\chi_1} - \frac{\chi_6}{\chi_2} < 0$. Only near the upper edge of the $2\sigma$ band on the high-temperature side does this difference become positive, thereby satisfying the complete ordering. For $T<T_{C,L}$, no consistent ordering is observed within either the $1\sigma$ or $2\sigma$ regions, although a reversed ordering, $\frac{\chi_5}{\chi_1} > \frac{\chi_4}{\chi_2} > \frac{\chi_3}{\chi_1}$, appears over a temperature interval near the lower edge of the $1\sigma$ band. These results demonstrate that the hierarchy among susceptibility ratios is not robust within the critical region and is only partially realized over a limited temperature interval above $T_{C,L}$. This lack of a robust, extended hierarchy in the $q=2$ state Potts model suggests that such cumulant-ratio inequalities may not be generic consequences of continuous phase transitions in finite systems.

\begin{figure*}[htbp!]
    \centering
    \includegraphics[width=0.325\linewidth]{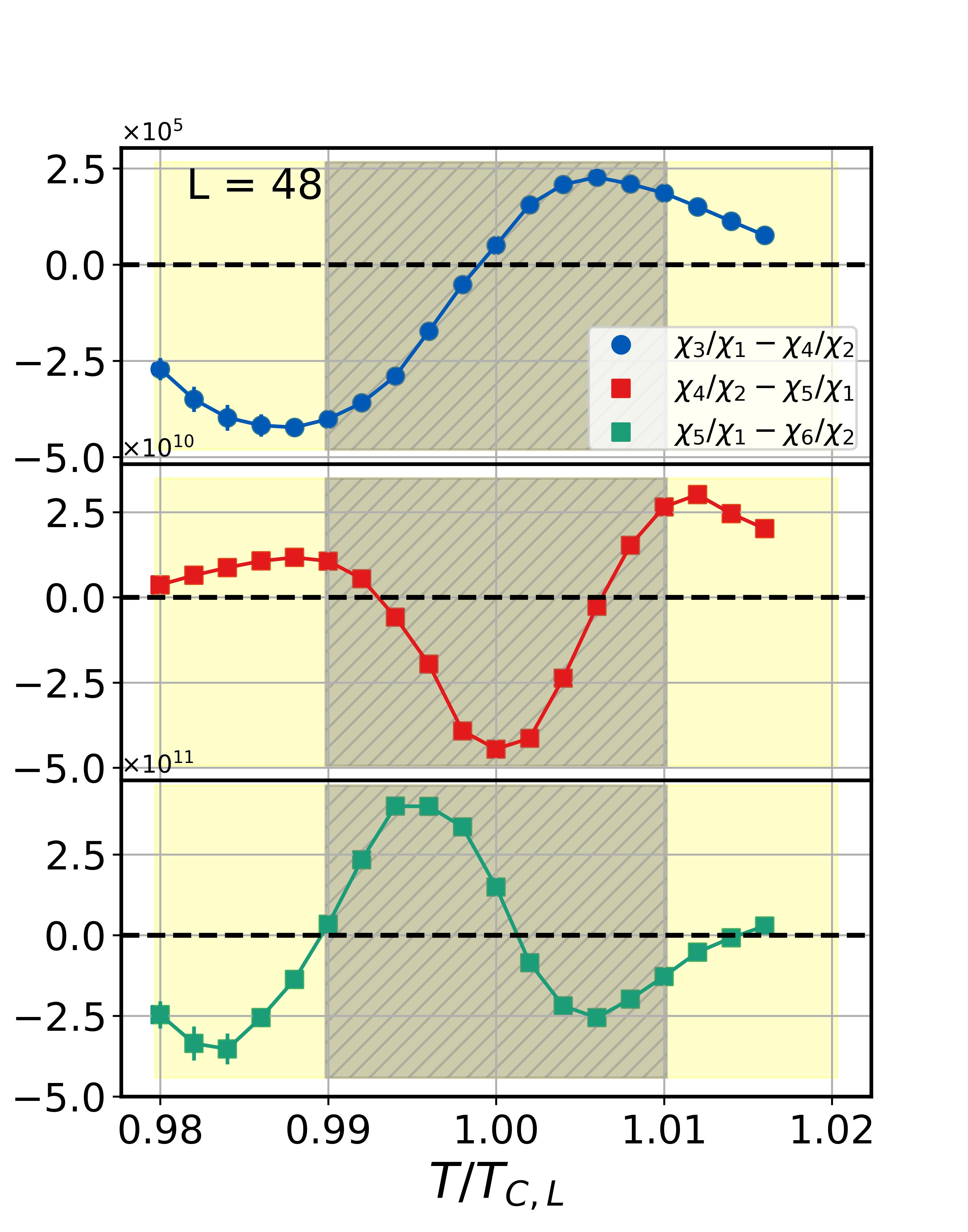}
    \includegraphics[width=0.325\linewidth]{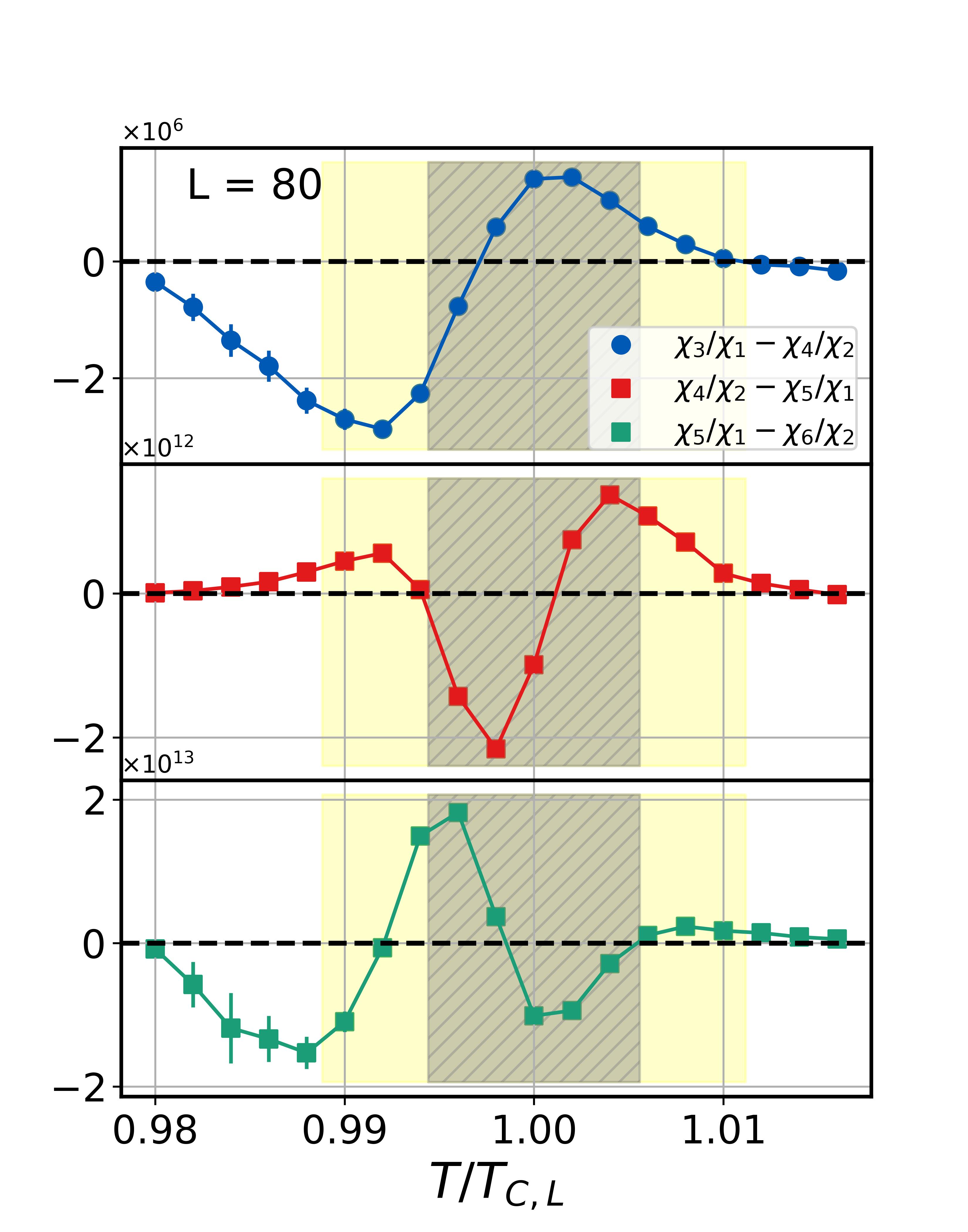}
    \includegraphics[width=0.325\linewidth]{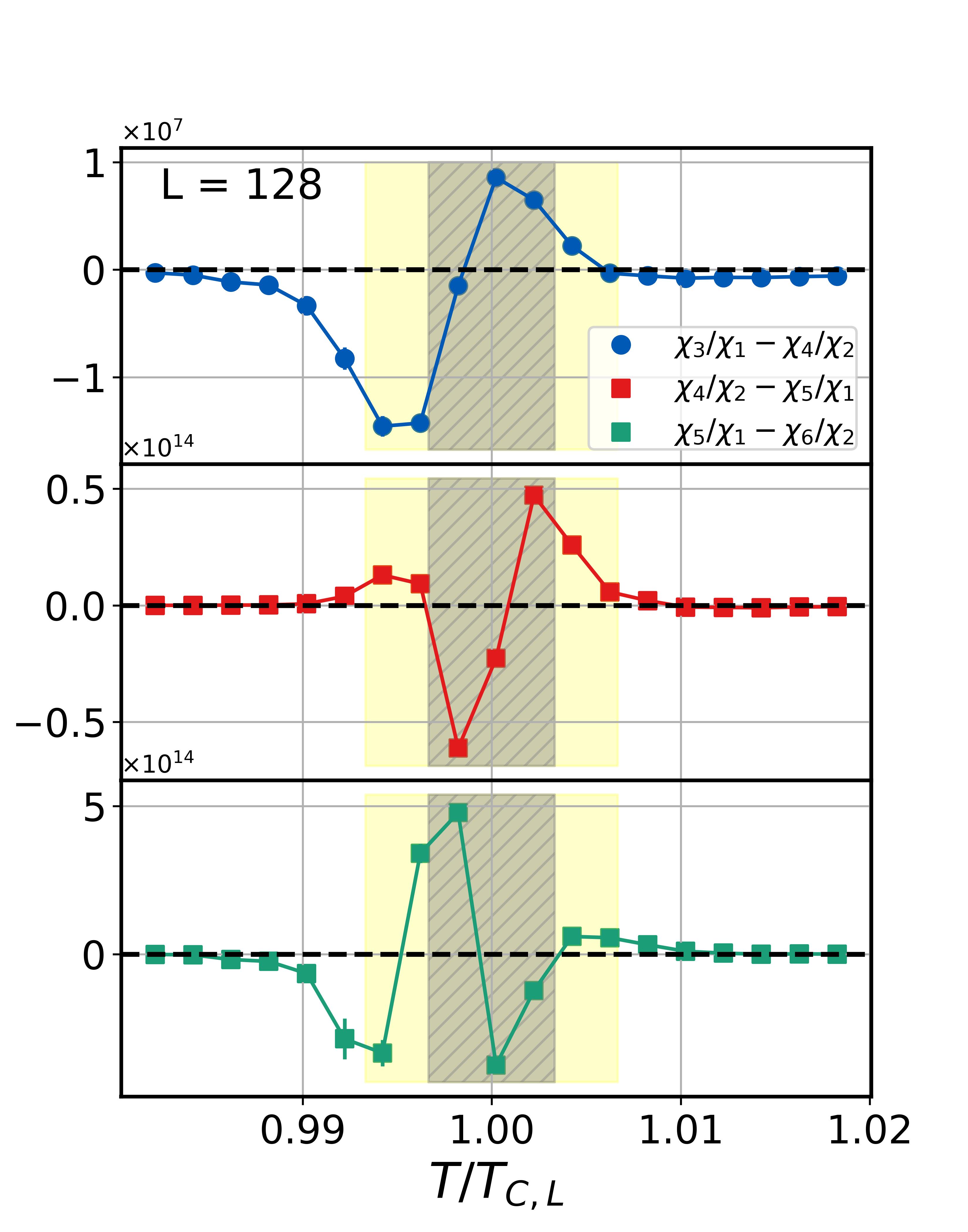}  
    \caption{The difference among ratios of the susceptibilities, ($\chi_3/\chi_1-\chi_4/\chi_2$), ($\chi_4/\chi_2-\chi_5/\chi_1$) and ($\chi_5/\chi_1-\chi_6/\chi_2$) for 2d $q = 3$ state Potts model with lattice sizes, L = 48 (left plot), 80 (middle plot), and 128 (right plot) are shown as a function of scaled temperature, $T/T_{C,L}$. The yellow shaded band indicates the critical region defined by $T_{C,L} \pm 2\sigma_{C,L}$, while the grey hatched region highlights the narrower interval $T_{C,L} \pm \sigma_{C,L}$.}
    \label{fig:potts_ho_ratios}
\end{figure*}

The differences among susceptibility ratios for the $q=3$ state Potts model are shown in Fig.~\ref{fig:potts_ho_ratios} for lattice sizes $L=48$, 80, and 128 (The comparison of individual ratios can be found in Fig.~\ref{fig:potts_ho_ratios_app}). Consistent with the $q=2$ case, the ordering $\frac{\chi_5}{\chi_1} < \frac{\chi_4}{\chi_2} < \frac{\chi_3}{\chi_1}$ emerges on the high-temperature side within the 1$\sigma$ region, setting in at $T \gtrsim T_a$, with $T_a \approx 1.008T_{C,L}$ for $L=48$ and $T_a \approx 1.002T_{C,L}$ for $L=80$ and 128. While this ordering persists on the high-temperature side within the 2$\sigma$ region for $L=48$ and 80, it is restricted to a narrow interval, $1.002T_{C,L} \lesssim T_a \lesssim 1.004T_{C,L}$, for $L=128$, since $\frac{\chi_3}{\chi_1}-\frac{\chi_4}{\chi_2}$ becomes negative for $T \gtrsim 1.006T_{C,L}$ even within the $2\sigma$ region. The hierarchy involving $\frac{\chi_6}{\chi_2}$ is not satisfied within the 1$\sigma$ region, as $\frac{\chi_5}{\chi_1} - \frac{\chi_6}{\chi_2}$ carries the opposite sign to the other two differences across all lattice sizes. Only within a narrow temperature interval on the high-temperature side in the $2\sigma$ region do all three differences become simultaneously positive, indicating a complete ordering. This interval shrinks with increasing lattice size, reducing from $1.004T_{C,L} \lesssim T \lesssim 1.008T_{C,L}$ for $L=80$ to a narrow region around $T \approx 1.004T_{C,L}$ for $L=128$. On the low-temperature side ($T < T_{C,L}$), no consistent hierarchy is observed within either the $1\sigma$ or $2\sigma$ regions. A partial reversed ordering, $\frac{\chi_5}{\chi_1} > \frac{\chi_4}{\chi_2} > \frac{\chi_3}{\chi_1}$, appears only over a narrow interval within the $1\sigma$ band, which shrinks with increasing lattice size and does not extend into the $2\sigma$ region.

Taken together, the behavior observed for both $q=2$ and $q=3$ models indicates that a partial reversal of the ordering up to $\frac{\chi_5}{\chi_1}$, i.e., $\frac{\chi_5}{\chi_1} > \frac{\chi_4}{\chi_2} > \frac{\chi_3}{\chi_1}$ occurs below $T_{C,L}$, while a complete reversal remains absent in both cases. An important point to note is that regions of inequality exclusively lie in the critical region, which itself shrinks in the infinite volume limit. This suggests that for infinite system size, we may not be able to establish any inequalities for the susceptibility ratios of magnetization in the spin models considered, and any observed orderings are a finite size effect. However, since the studied inequalities are relevant to LQCD and experimental heavy-ion collisions, both of which deal with systems with finite volume (moderate $L/\xi$), this work holds significant importance. Our preliminary determinations of the correlation lengths suggest that the ratio of $L$ to the correlation length ($\xi$), denoted as $L/\xi$, ranges between approximately 9.1 and 13.3 for the lattice systems studied in this work. Notably, systems associated with LQCD studies and experimental heavy-ion collisions exhibit $L/\xi$ values lower than those obtained in our study, indicating that finite size effects also play a crucial role in shaping the observed behavior in those systems. It would be intriguing to explore whether the volume dependency observed for the inequalities in spin models also holds true for the produced medium in relativistic heavy-ion collisions.

\section{\label{sec:Conclusion} Conclusions}
A specific hierarchy of net-baryon number susceptibility ratios near the quark-hadron transition temperature at small $\mu_\mathrm{B}$, namely $\frac{\chi_6}{\chi_2} < \frac{\chi_5}{\chi_1} < \frac{\chi_4}{\chi_2} < \frac{\chi_3}{\chi_1}$, has been proposed in the literature~\cite{PhysRevD.101.074502,PhysRevD.96.074510,PhysRevD.104.094047}. The STAR experiment tested these predictions for net-proton number cumulant ratios in Au+Au collisions at various center-of-mass energies. Cumulant ratios from 7.7 to 200 GeV exhibited the predicted hierarchy, while a reverse ordering was observed at 3 GeV. If these inequalities are universally applicable for any system undergoing phase transition, is what we tried to understand.

Although the QCD critical point belongs to the three-dimensional Ising universality class, the qualitative behavior of fluctuations near criticality and their sensitivity to finite system size can be explored in simpler statistical systems such as spin models. In this study, we investigated higher-order susceptibilities of magnetization and their ratios in the critical regions of 2d two and three-state Potts models. Utilizing the Wolff cluster algorithm to mitigate critical slowing down effects, simulations were conducted. We employed finite size scaling techniques to extract scaling exponents associated with higher-order susceptibilities, extending up to the sixth order. Additionally, we estimated these scaling exponents using the known critical exponents of the model, specifically the $\nu$ and $\gamma$ exponents. The obtained values from both methods exhibited a good agreement, indicating that the behavior of higher-order susceptibilities aligns well with our understanding of second-order phase transitions in the spin models.

We do not observe a consistent ordering of susceptibility ratios either above or below the pseudo-critical temperature. A truncated hierarchy, $\frac{\chi_5}{\chi_1} < \frac{\chi_4}{\chi_2} < \frac{\chi_3}{\chi_1}$, appears only on the high-temperature side within a narrow temperature interval in the $1\sigma$ band of the critical region for both $q=2$ and $q=3$ state Potts models. The hierarchy involving $\frac{\chi_6}{\chi_2}$ is not satisfied within the $1\sigma$ region. A complete ordering of the susceptibility ratios is realized for both models only in an even narrower temperature interval on the high-temperature side outside the $1\sigma$ region, near the upper edge of the $2\sigma$ band; this interval shrinks with increasing lattice size. The statistical robustness of this feature is assessed using resampling-based uncertainty estimates. However, given the limited width of the temperature interval in which this ordering is observed, it should be interpreted as a finite-size effect whose visibility depends sensitively on statistical precision and analysis details. 

Below the pseudo-critical temperature, a complete reversal of this ordering is not observed in either model; only a partial reversal, $\frac{\chi_5}{\chi_1} > \frac{\chi_4}{\chi_2} > \frac{\chi_3}{\chi_1}$, is seen within a small temperature range. Furthermore, the temperature region over which meaningful comparisons between susceptibility ratios can be made diminishes notably as the lattice size (or ``lattice volume'') increases, and is expected to shrink to a point in the infinite volume limit. These results therefore demonstrate that while critical scaling is a universal feature of finite systems near second-order phase transitions, the ordering of higher-order cumulant ratios is not, highlighting the limitations of universality arguments for such observables. Instead, such ordering emerges only within narrow temperature intervals and weakens with increasing system size, indicating a strong sensitivity to model details and finite-size effects. Consequently, the orderings observed in the susceptibility ratios can be attributed to finite size effects in spin models.

The 2d spin models considered here have only a second order phase transition.  A more realistic analog to the QCD phase diagram is provided by the 3d $q=3$ state Potts model with a small, non-vanishing magnetic field $h$, where the first-order transition at $h=0$ weakens into a critical endpoint as $h \to h_c$ (critical magnetic field). This establishes a qualitative correspondence between the $(T,h)$ plane---encompassing first-order, critical, and crossover regions---and the QCD $(T,\mu_\mathrm{B})$ plane. In such a framework, the behavior of higher-order cumulants may depend on the trajectory along which they are evaluated. For instance, following a path at fixed temperature $T_{fc}$ on the first-order line by varying only $h$ does not necessarily pass through or probe the vicinity of the critical point, which may thereby affect the ordering of cumulant ratios. This suggests that the absence of a robust hierarchy in the present study may be influenced by the specific path (fixed $h=0$) in parameter space, indicating that the ordering of cumulant ratios may not be a universal feature even within a given model. It would therefore be of interest to investigate how the hierarchy of cumulants evolves across the different regions of the 3d Potts $(T,h)$ plane. In addition, for a meaningful comparison with experimental measurements in heavy-ion collisions, global conservation laws must be taken into account. Baryon number conservation, which holds event-by-event, can modify higher-order net-baryon cumulants in a non-linear manner. Also, since measurements in experiment are performed within a limited kinematic acceptance, the observed cumulant ratios depend on the fraction of the system that is probed, and the resulting hierarchy may be altered. Such effects are not included in the present spin-model study and should be considered for a more direct comparison with experimental data.

\ack
We acknowledge the support of the Department of Atomic Energy (DAE) and the Department of Science and Technology (DST), Government of India. RVG expresses gratitude for the financial support from DAE through the Raja Ramanna Fellowship and the hospitality of the Department of Physics at IISER Bhopal. BM and SS acknowledge the financial support of DAE and DST. We also acknowledge the use of Garuda HPC facility at the School of Physical Sciences, NISER. SS would like to thank V K S Kashyap for technical help in running simulations, as well as Ashish Pandav and Debasish Mallick for insightful discussions regarding STAR results.

\section{Data availability statement}
Our data that underlie the findings are presented graphically within the article and are available in tabular format in the Supplementary Material.

\appendix

\newpage
\section{Ratio of cumulants}

\begin{figure*}[h]
  \centering
    \includegraphics[width=0.325\linewidth]{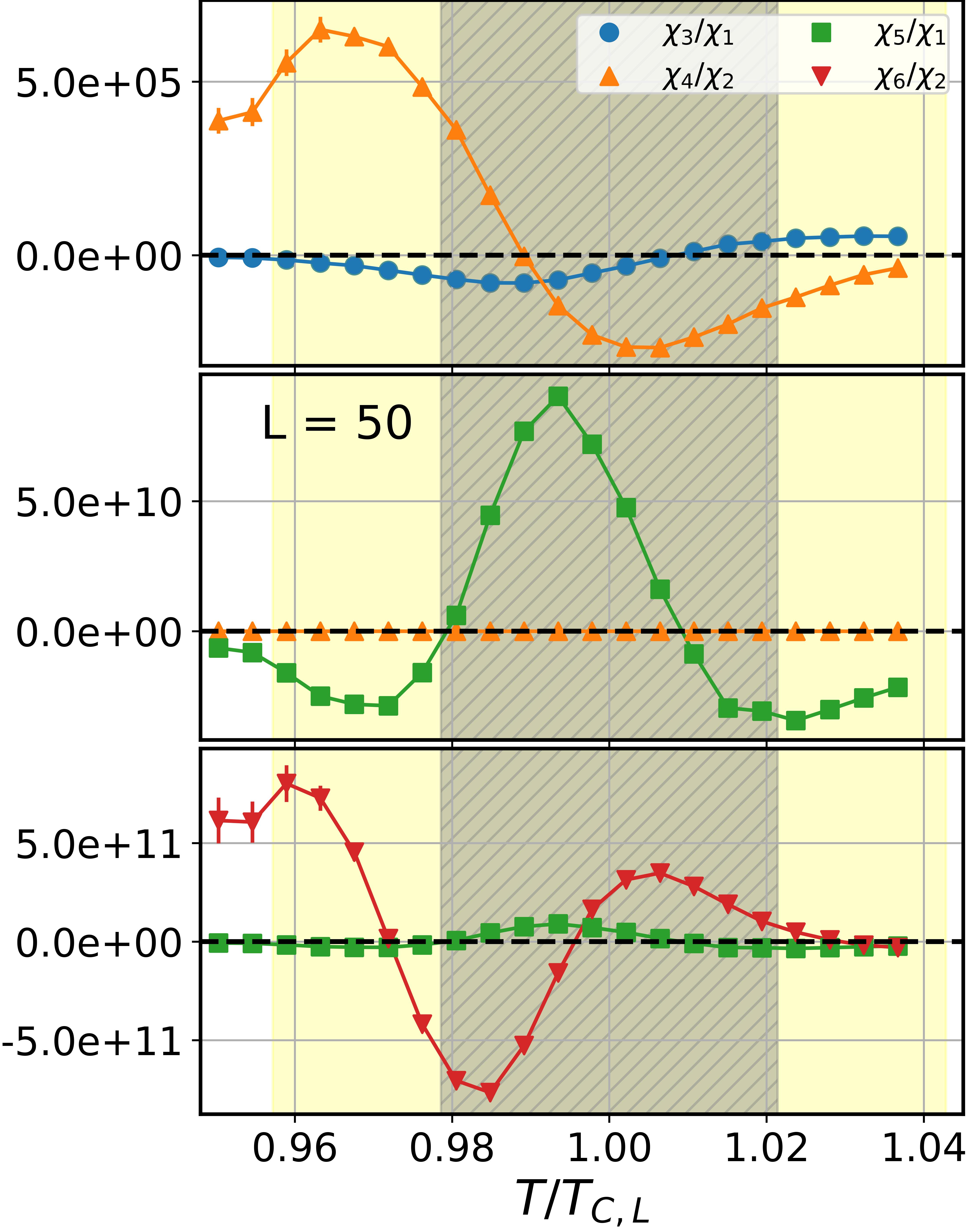}
    \includegraphics[width=0.325\linewidth]{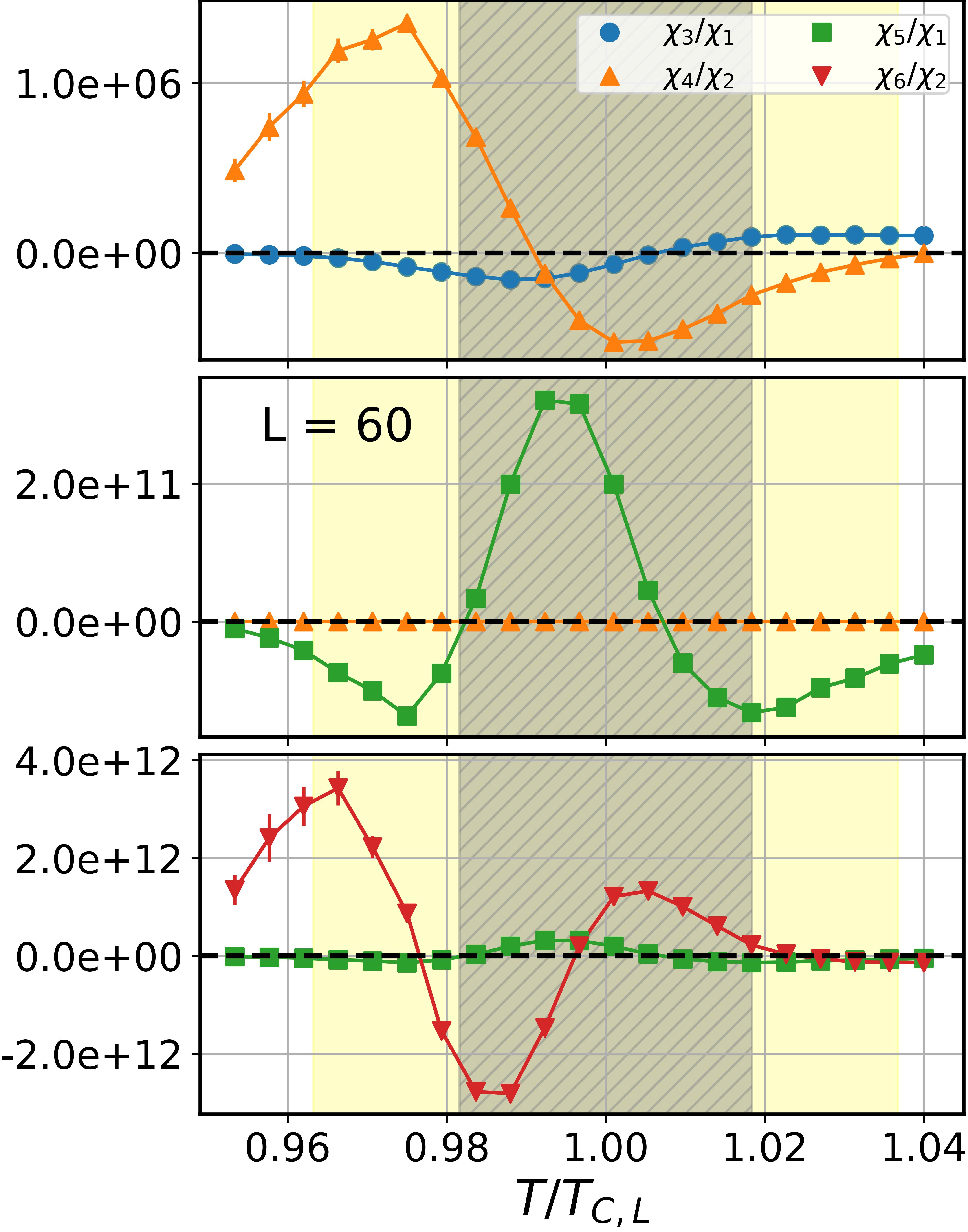}
    \includegraphics[width=0.325\linewidth]{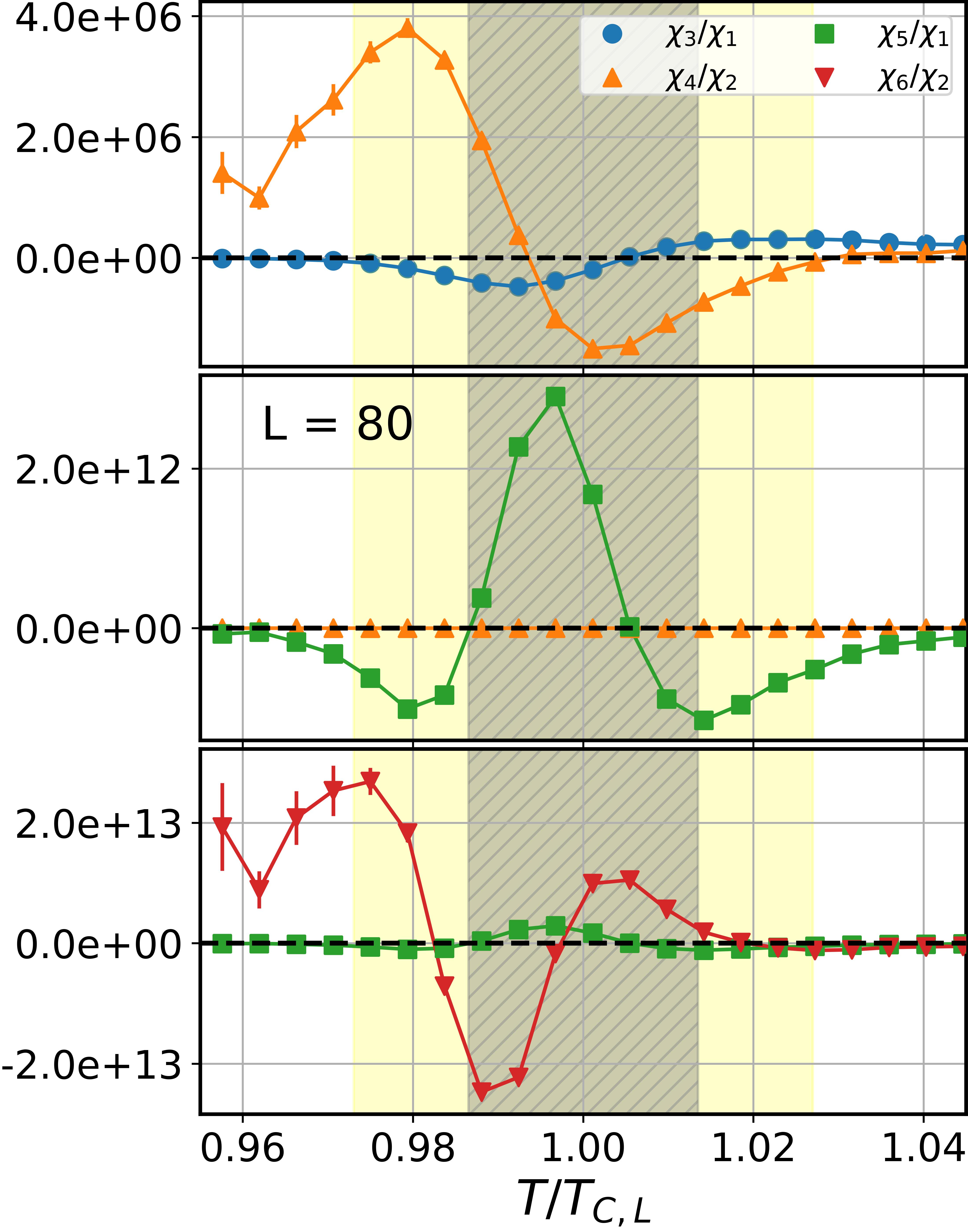} 
    \caption{Ratios of the susceptibilities, $\chi_3/\chi_1$, $\chi_4/\chi_2$, $\chi_5/\chi_1$ and $\chi_6/\chi_2$ for 2d $q = 2$ state Potts model with lattice sizes, L = 50 (left plot), 60 (middle plot), and 80 (right plot) are shown as a function of scaled temperature, $T/T_{C,L}$. The yellow shaded band indicates the critical region defined by $T_{C,L} \pm 2\sigma_{C,L}$, while the grey hatched region highlights the narrower interval $T_{C,L} \pm \sigma_{C,L}$.}
    \label{fig:Ising_ratios_app}
\end{figure*}
\begin{figure*}[h]
    \centering
  \includegraphics[width=0.325\linewidth]{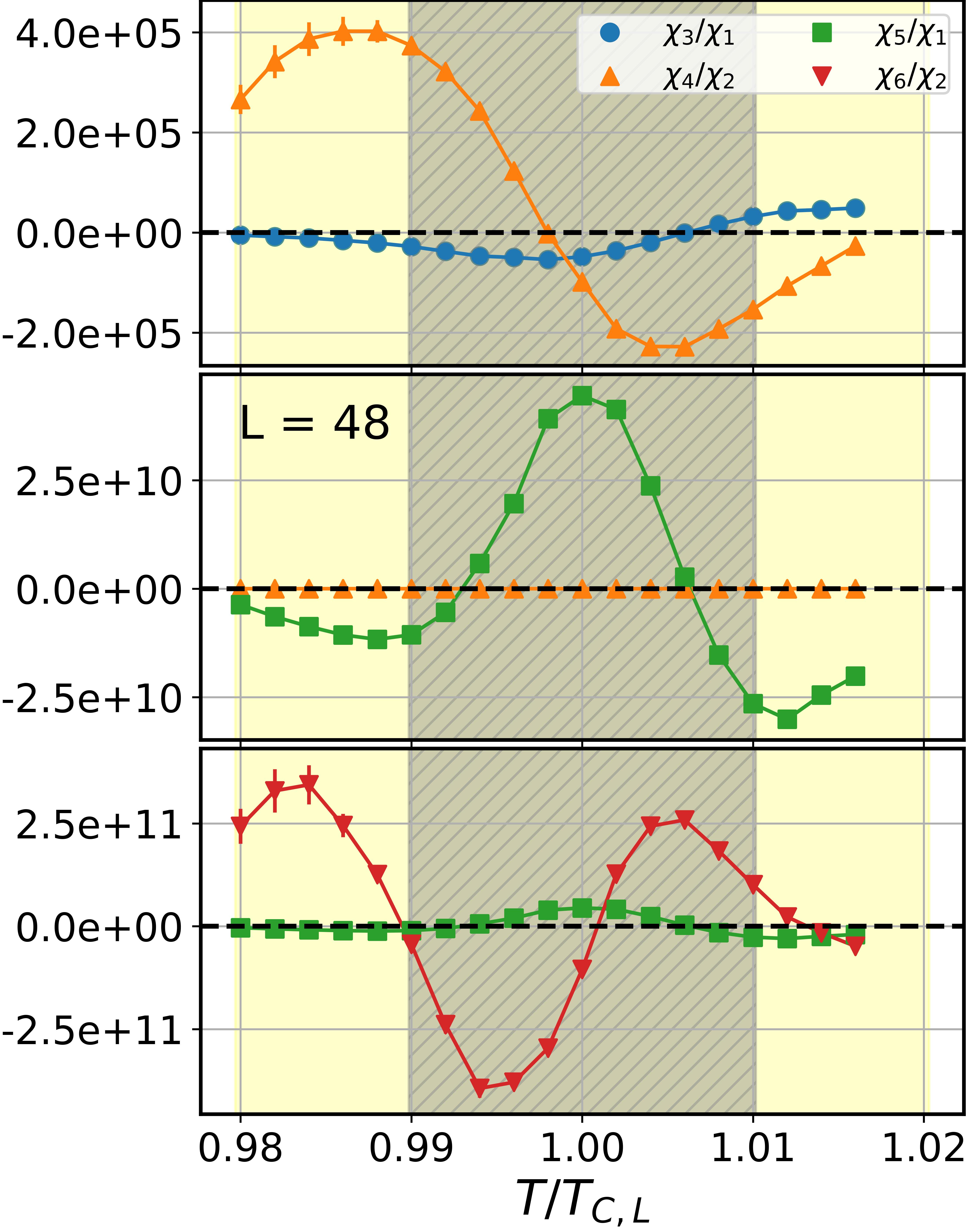}
  \includegraphics[width=0.325\linewidth]{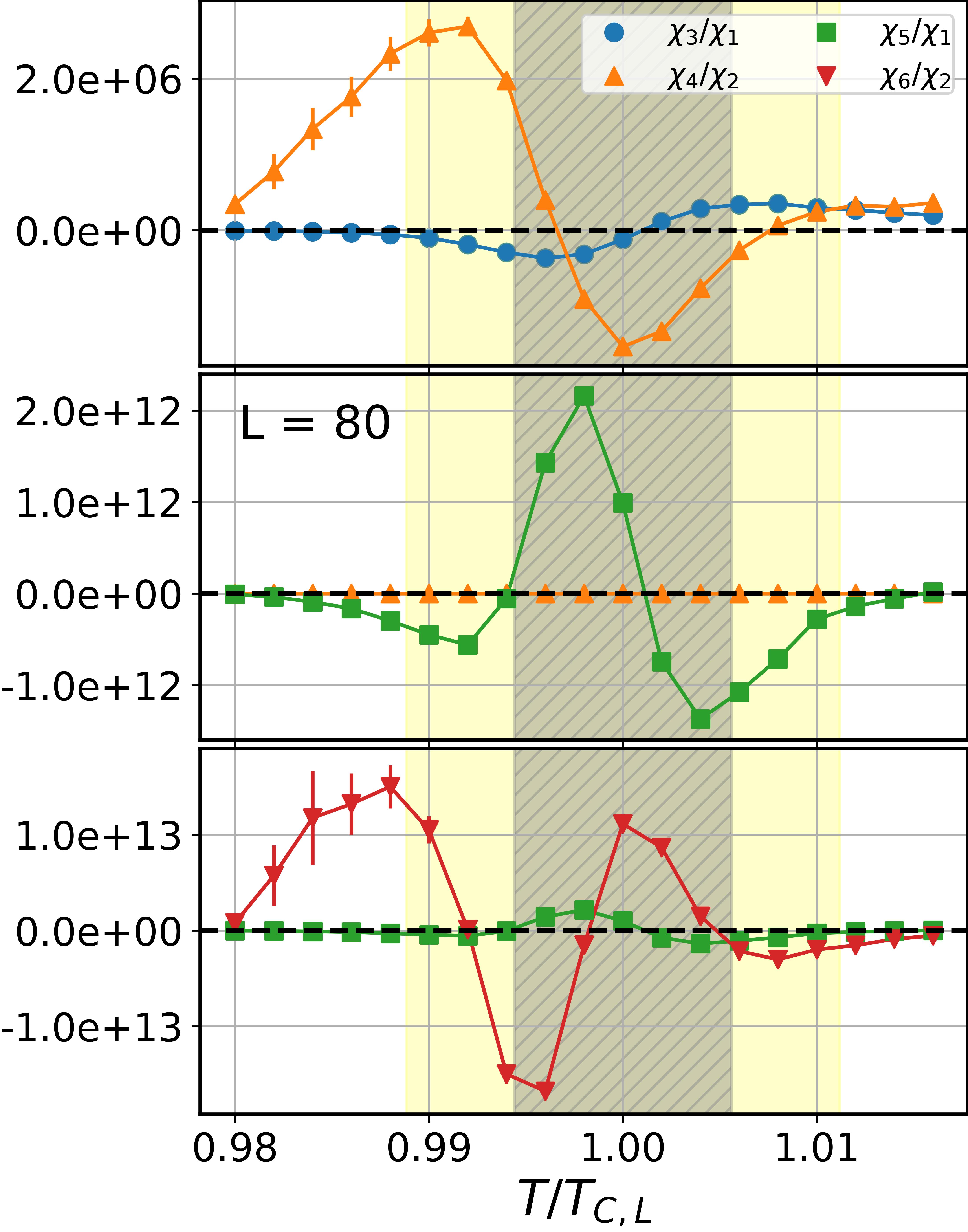}
  \includegraphics[width=0.325\linewidth]{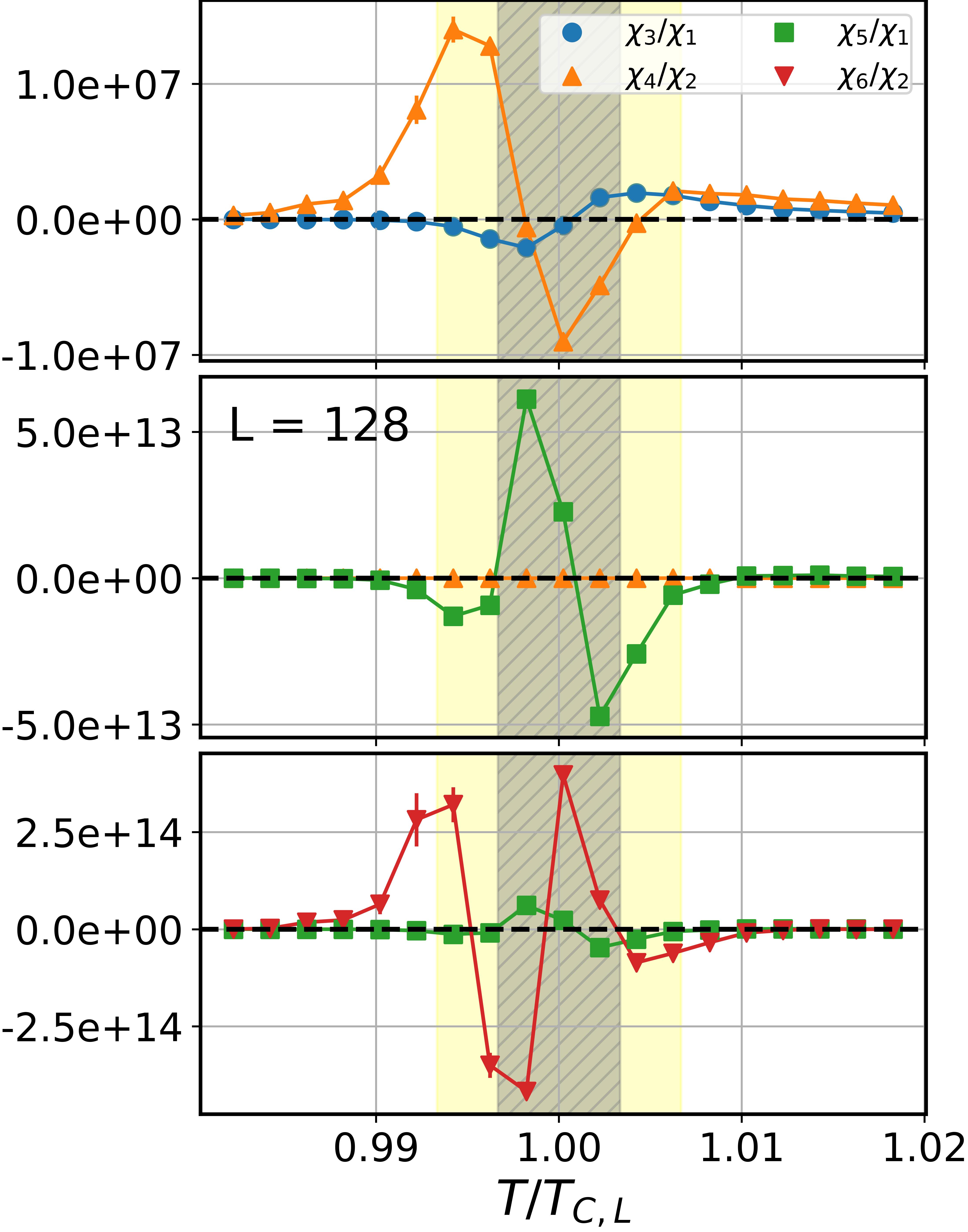}  
    \caption{Ratios of the susceptibilities, $\chi_3/\chi_1$, $\chi_4/\chi_2$, $\chi_5/\chi_1$ and $\chi_6/\chi_2$ for 2d $q = 3$ state Potts model with lattice sizes, L = 48 (left plot), 80 (middle plot), and 128 (right plot) are shown as a function of scaled temperature, $T/T_{C,L}$. The yellow shaded band indicates the critical region defined by $T_{C,L} \pm 2\sigma_{C,L}$, while the grey hatched region highlights the narrower interval $T_{C,L} \pm \sigma_{C,L}$. }
    \label{fig:potts_ho_ratios_app}
\end{figure*}

\section*{References}
\bibliographystyle{unsrt.bst}
\bibliography{references.bib}

\end{document}